\documentclass[]{pasj02}

\jyear{2026}
\Received{$\langle$reception date$\rangle$}
\Accepted{$\langle$acception date$\rangle$}
\Published{$\langle$publication date$\rangle$}

\usepackage{comment}
\usepackage{caption}
\usepackage{natbib,bm}
\usepackage{enumitem}

\usepackage{tabularx}
\usepackage{makecell}
\usepackage{booktabs, multirow, array}

\usepackage[switch,mathlines]{lineno}

\newcommand{\beq}{\begin{equation}}
\newcommand{\eeq}{\end{equation}}
\newcommand{\beqn}{\begin{eqnarray}}
\newcommand{\eeqn}{\end{eqnarray}}
\newcommand{\mum}{$\rm{\mu m}$}

\newcommand{\eqref}[1]{(\ref{#1})}

\graphicspath{{./}{figure/}}

\title{Constraints on the Crystallinity of Water Ice in Planet-forming Disks from Infrared Scattered-Light Spectra}

\author{Kanon \textsc{Nakazawa}\altaffilmark{1}}
\altaffiltext{1}{Department of General Systems Studies, The University of Tokyo, Tokyo 153-8902, Japan}

\author{Ryo \textsc{Tazaki}\altaffilmark{1}}

\email{kanon-nakazawa@g.ecc.u-tokyo.ac.jp}

\KeyWords{infrared: planetary systems --- Kuiper belt: general --- protoplanetary disks} 

\maketitle

\begin{document}
\begin{abstract}
    The crystallinity of water ice not only records the thermal history experienced by an astronomical body, but also affects the composition of forming planets by controlling the trapping of volatile materials in amorphous ice and their subsequent transport. An additional structure within the 3~\mum\ water-ice absorption band, known as the Fresnel feature, may serve as a diagnostic of ice crystallinity. Recent observations with the James Webb Space Telescope have detected a Fresnel peak in a debris disk and in Trans-Neptunian Objects (TNOs).
    Here, we propose a portable expression that translates the observed Fresnel peak strength into the degree of crystallinity of icy grains in debris disks. Our formula targets scattered light at around 90$^{\circ}$ angles, which are easily accessible for spatially resolved debris disks regardless of the inclination angle. Applying this expression, we derive the degree of crystallinity of a debris disk around HD 181327 to be 10--20\%. We also study the Fresnel feature in protoplanetary disks and find that it is generally weaker than in debris disks even for the same crystallinity. 
    We then analyzed a scattered light spectrum of the protoplanetary disk around d216-0939, which shows a weak crystalline feature, and inferred a crystallinity of $\sim$50\%.  We conclude that the Fresnel feature is a reliable observational tracer for ice crystallinity, and future near-IR spectroscopic observations will be crucial to elucidate the crystalline ice evolution.
\end{abstract}
%\pagewiselinenumbers
%%%%%%%%%%%%%%%%%%%%%%%%%%%%%%%%%%%
\begin{figure*}[t]
\centering
\includegraphics[width= 0.8\hsize]{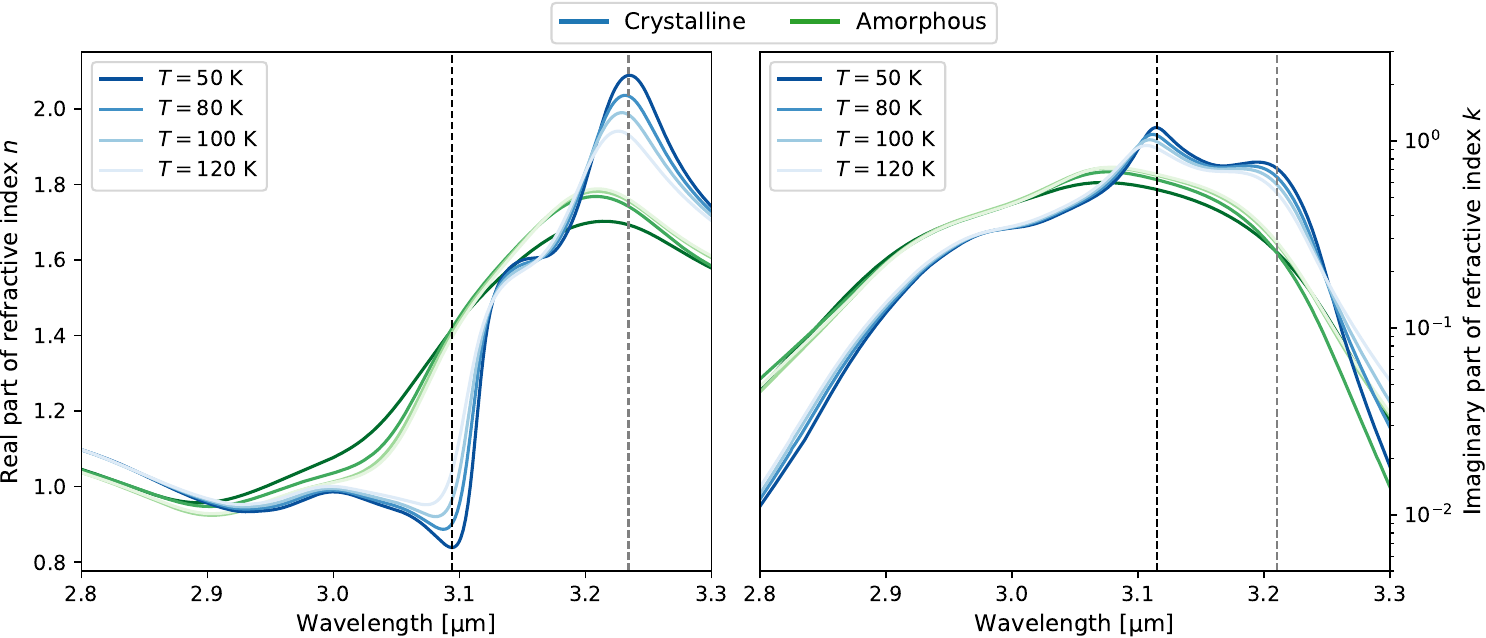}
\caption{Wavelength dependence of the real and imaginary parts of the complex refractive indices of crystalline and amorphous water ice measured by \citet{2009ApJ...701.1347M}. Blue lines correspond to crystalline ice and green lines to amorphous ice. The saturation of the colors indicates the ice measurement temperature. The black and gray vertical lines highlight wavelengths at which crystalline ice at $T=50$~K exhibits characteristic peaks: $n$ at 3.094 and 3.234 \mum, and $k$ at 3.115 and 3.210 \mum. {Alt text: Two line graphs showing the complex refractive indices of water ice.}}
\label{fig:refractive_index}
\end{figure*}
%%%%%%%%%%%%%%%%%%%%%%%%%%%%%%%%%%

%%%%%%%%%%%%%%%%%%%%%%%%%%%%%%%%%%%
\begin{figure*}[t]
\includegraphics[width= \hsize]{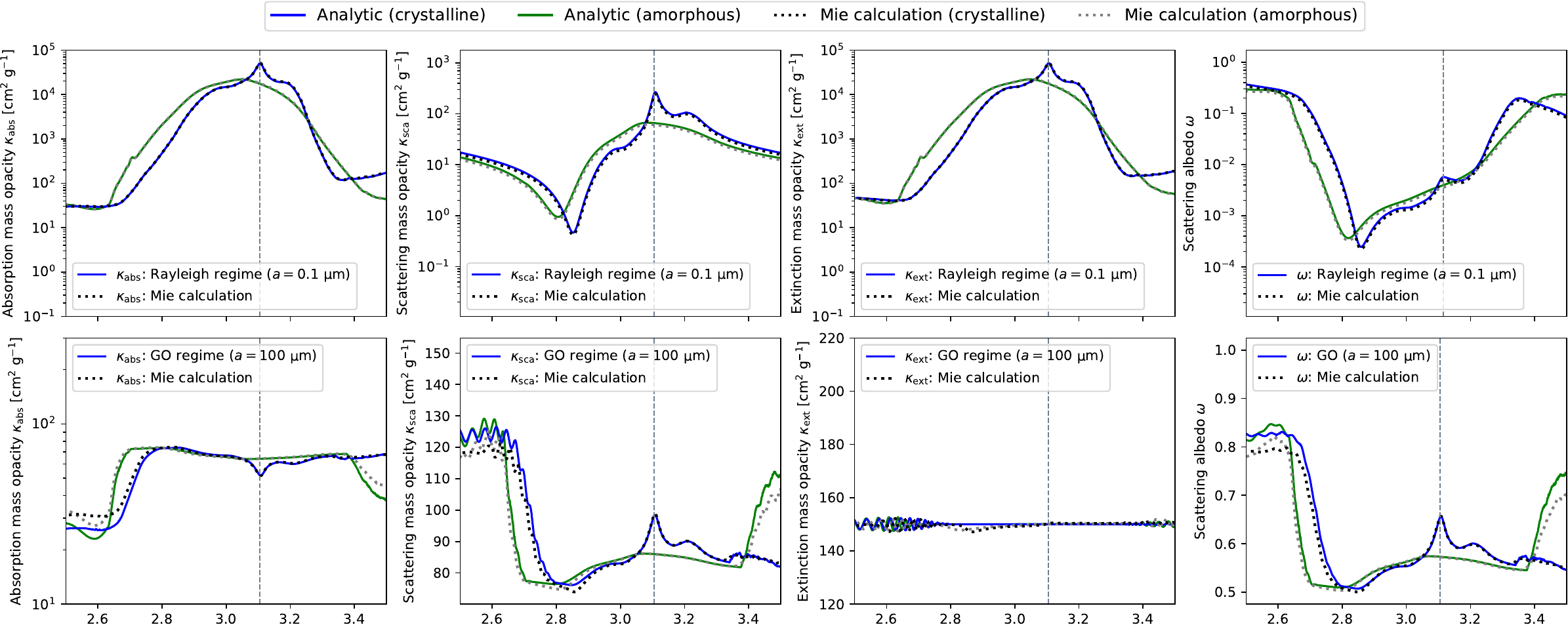}
\caption{Optical properties of crystalline and amorphous water ice obtained from the analytic expressions in equations~\eqref{kappa_abs_Rayleigh}--\eqref{kappa_sca_GO} (blue and green solid lines) and those obtained from Mie calculations performed with \texttt{OpTool} (black and gray dotted lines). The top row shows the Rayleigh-regime case, and the bottom row shows the geometric-optics-regime case. From left to right, the panels present the absorption mass opacity, scattering mass opacity, extinction mass opacity, and the single-scattering albedo. The light gray vertical line is a guide to the eye marking 3.105~\mum, where the Fresnel feature appears. {Alt text: Line graphs showing the optical properties of water ice.}}
\label{fig:optical_property_Mie}
\end{figure*}
%%%%%%%%%%%%%%%%%%%%%%%%%%%%%%%%%%

\section{Introduction}
Icy dust is a fundamental constituent of protoplanetary disks, and water ice, the most abundant ice species, plays a critical role in multiple aspects of planet formation. By coating dust‐grain surfaces, ices provide sites for chemical reactions and promote the formation of larger compounds, including organic molecules \citep{2009ARA&A..47..427H}. Radial drift of water ice can deliver life-essential H$_2$O to forming habitable planets in the inner disk \citep{2016A&A...589A..15S, 2019A&A...624A..28I}. This transport can extend to a variety of volatile species through molecular entrapment in amorphous water ice, potentially carrying hyper-volatiles such as CO and N$_2$ into the inner regions as well \citep{1985Icar...63..317B, 1988PhRvB..38.7749B, 2007Icar..190..655B, 2004MNRAS.354.1133C, 2019ApJ...875....9M, 2023ApJ...955....5S}. In addition, ice modifies the stickiness of grains, thereby influencing the efficiency of dust growth \citep{1997ApJ...480..647D, 2009ApJ...702.1490W, 2013A&A...559A..62W}. Despite its significance for planet formation, constraints on the abundance and phase of ice within astronomical bodies remain limited.

The advent of the James Webb Space Telescope (JWST), with its high spectral and spatial resolution, offers an unprecedented opportunity to conduct integral field unit observation of the icy dust that composes protoplanetary disks and debris disks. The presence of water ice can be identified from its near- to mid-infrared scattering signatures, most prominently the H$_2$O feature near 3 $\rm{\mu m}$ arising from the O-H stretching. Currently, JWST has identified H$_2$O ice in several edge-on protoplanetary disks \citep[e.g., ][]{2023A&A...679A.138S, 2025A&A...697A..53P, 2025A&A...698A...8D, 2026arXiv260318163B}, in the face-on debris disk HD 181327 \citep{2025Natur.641..608X}, and among Trans-Neptunian Objects (TNOs) in the Solar System \citep{2025NatAs...9..245L, 2025NatAs...9..230P}. In particular, the outer ring of HD 181327 and KBOs exhibit the 3.1 \mum\ Fresnel peak characteristic of crystalline ice \citep{2006Sci...311.1425B}.

The degree of ice crystallinity not only affects the entrapment efficiency of volatile molecules but also provides a diagnostic of the thermal histories of celestial bodies. In general, water ice occurs in two structural types: amorphous and crystalline. Under the cryogenic environment ($\lesssim 100$ K) typically in interstellar space or outer region of disks, water ice form amorphous ice lacking large-scale regularity in molecular arrangements \citep{1994Sci...265..753J}. Upon heating to around 140 K, amorphous ice undergoes a phase transition to cubic or hexagonal crystalline ice \citep{1990JCrGr..99.1220K, 2024come.book..823P}. This transition is irreversible with respect to subsequent cooling; re-amorphization requires sustained irradiation by ultraviolet photons, energetic electrons, or ions \citep{1990Natur.344..134K, 2004JGRE..109.1012H}. Consequently, the detection of crystalline ice in debris disks and on KBOs suggests that their surfaces are mantled by fresh grains that preserve the record of past heating events.

Given the importance of the abundance and physical state of ice for planet formation, it is essential to predict their spectral signatures systematically through theoretical modeling. Based on laboratory measurements of the complex refractive indices of amorphous and crystalline water ice, \citet{2024MNRAS.533.2801K} have computed the optical properties of dust grains and quantified how parameters such as grain size and ice abundance/phase influence observations of debris disks around stars of various spectral types. They have shown that features near 2.7 and 3.3 $\rm{\mu m}$ depend on both ice content and scattering geometry, exhibiting the strongest signatures under backscattering conditions. They have also demonstrated that the 3.1 $\rm{\mu m}$ Fresnel peak serves as a diagnostic of the water ice phase and identified JWST filter configurations that are effective for constraining grain size and ice content.

A key next step is to predict scattered-light spectra for bodies composed of dust with a range of grain sizes, ice fractions, and degrees of crystallinity, and to quantify how crystalline-ice features depend on these parameters. Since current observational interpretation often relies on radiative-transfer simulations with Monte Carlo sweeps over dust physical parameters and composition to identify plausible values, it is critical to develop a formulation that estimates ice crystallinity directly from the observed spectrum alone. To this end, we perform radiative-transfer simulations and imaging of debris and protoplanetary disks spanning broad ranges in grain size, ice content, and crystallinity. We then quantify how these parameters affect the crystalline-ice features in scattered light. We apply the resulting diagnostics to observed spectra to infer crystallinity and discuss the thermal histories and origins of the grains. This paper is organized as follows: In Section \ref{sec:fresnel_feature}, prior to the radiative-transfer simulations of planet-forming disks, we analyze the optical properties of crystalline and amorphous water ice in the Rayleigh and geometric-optics regimes and provide an overview of why crystalline ice exhibits the Fresnel feature. In Section \ref{sec:simulation_method}, we describe our simulation methods. We present the results of scattered-light imaging for debris disks in Section \ref{sec:debri_res}. We then present the scattered-/transmitted-light imaging results for face-on and edge-on protoplanetary disks in Section \ref{sec:ppds_res}. In Section \ref{sec:discussions}, we discuss several implications, including estimates of ice crystallinity in real systems based on comparisons between observed Fresnel features and our models. Section \ref{sec:conclusions} summarizes our conclusions.

\section{Analytic overview of the H$_2$O Fresnel feature}
\label{sec:fresnel_feature}

The H$_2$O Fresnel feature in the near-infrared arises from differences in the complex refractive indices of crystalline and amorphous water ice and provides a key diagnostic of the ice phase that is largely insensitive to the characteristic grain size. \citet{2008PASJ...60..557I}, using an approximate treatment of the radiative transfer equation under the assumption of isotropic scattering, showed that for face-on disks the scattered-light surface brightness scales with the scattering mass opacity, $\kappa_{\rm sca}$, in optically thin disks, whereas it scales with the scattering albedo, $\omega = \kappa_{\rm sca}/(\kappa_{\rm abs} + \kappa_{\rm sca})$, in optically thick disks, where $\kappa_{\rm abs}$ is the absorption mass opacity. Accordingly, before turning to scattered-light imaging of disks, we review the complex refractive indices of crystalline and amorphous water ice and show how their differences produce the characteristic Fresnel feature around 3.1 \mum\ through the behavior of $\kappa_{\rm sca}$ and $\omega$, using analytic expressions in two grain-size limits as well as Mie calculations.

For simplicity, we consider scattering by compact, homogeneous spherical grains. The scattering properties are characterized by the size parameter,
\begin{equation}
\label{size_parameter}
    x = \frac{2\pi a}{\lambda},
\end{equation}
where $a$ is the grain radius and $\lambda$ is the wavelength, together with the complex refractive index of the material,
\begin{equation}
\label{refractive_index}
    m = n+ik.
\end{equation}
Figure~\ref{fig:refractive_index} shows the real and imaginary parts of the complex refractive indices of crystalline and amorphous ice measured in the laboratory \citep{2009ApJ...701.1347M}. Compared to amorphous ice, crystalline ice exhibits an additional local minimum in $n$ at 3.1~\mum\ and a higher peak near 3.25~\mum. The imaginary part $k$ also differs: crystalline ice shows two peaks at 3.1 \mum\ and 3.2~\mum, whereas amorphous ice has a simpler profile with a single local maximum near 3.1~\mum. These features display a modest temperature dependence, with the crystalline-ice signatures becoming more pronounced at lower temperatures.

To analyze the scattering behavior around 3.1~\mum, we consider the absorption and scattering mass opacities, $\kappa_{\rm abs}$ and $\kappa_{\rm sca}$, in two limiting regimes \citep{1983asls.book.....B}. 
Absorption and scattering by small grains ($x \ll 1$ and $|m|x \ll 1$) can be described in the Rayleigh regime, for which
\begin{equation}
\label{kappa_abs_Rayleigh}
    \kappa_{\rm abs}^{\rm Rayleigh} = \frac{\pi a^2}{m_{\rm d}} 4x{\rm Im}\left(\frac{m^2-1}{m^2+2}\right),
\end{equation}
and
\begin{equation}
\label{kappa_sca_Rayleigh}
    \kappa_{\rm sca}^{\rm Rayleigh} = \frac{8\pi a^2}{3m_{\rm d}}x^4\left|\frac{m^2-1}{m^2+2}\right|^2.
\end{equation}
where $m_{\rm d}$ is the grain mass.

In contrast, absorption and scattering by large, optically thick grains ($x \gg 1$ and $|m-1|x \gg 1$) can be described in the geometric-optics (GO) regime \footnote{Note that equations (\ref{kappa_abs_GO}, \ref{kappa_sca_GO}) are not valid for less absorbing grains (i.e., $k \ll 1$). For water ice, the geometric-optics approximation provides a good match to Mie theory in the 2.9--3.3~\mum\ range that includes the Fresnel feature, whereas at other wavelengths it is preferable to adopt alternative approximations such as anomalous diffraction theory \citep{1957lssp.book.....V}.},
\begin{equation}
\label{kappa_abs_GO}
    \kappa_{\rm abs}^{\rm GO} = \frac{\pi a^2}{m_{\rm d}}\int_{0}^{\pi/2} \{1-R(\theta_i)\}\sin2\theta_id\theta_i,
\end{equation}
and
\begin{equation}
\label{kappa_sca_GO}
    \kappa_{\rm sca}^{\rm GO} = \frac{\pi a^2}{m_{\rm d}}\int_{0}^{\pi/2} \{1+R(\theta_i)\}\sin2\theta_id\theta_i,
\end{equation}
where the reflectance $R$ is given by
\begin{equation}
\label{Reflectance}
    R(\theta_i) = \frac{1}{2}\left(\left|\frac{\cos\theta_t-m\cos\theta_i}{\cos\theta_t+m\cos\theta_i}\right|^2+
    \left|\frac{\cos\theta_i-m\cos\theta_t}{\cos\theta_i+m\cos\theta_t}\right|^2\right),
\end{equation}
and the incident angle $\theta_i$ and refracted angle $\theta_t$ satisfy Snell’s law,
\begin{equation}
\label{Snell}
    \sin\theta_t = \frac{\sin\theta_i}{m}.
\end{equation}
Figure~\ref{fig:optical_property_Mie} presents $\kappa_{\rm abs}$, $\kappa_{\rm sca}$, extinction mass opacity $\kappa_{\rm ext} = \kappa_{\rm abs} + \kappa_{\rm sca}$, and $\omega$ for $a=0.1$ \mum~ice computed from equations \eqref{kappa_abs_Rayleigh} and \eqref{kappa_sca_Rayleigh}, and for $a=100$ \mum~ice computed from equations \eqref{kappa_abs_GO} and \eqref{kappa_sca_GO}. An important property is that crystalline ice exhibits an upward convex peak at 3.105~\mum\ in both $\kappa_{\rm sca}$ and $\omega$, irrespective of particle size. This implies that, when crystalline ice is present, a Fresnel "peak" is expected to appear at 3.1~\mum\ in disk scattered light. At first glance, this may seem counterintuitive, because the Fresnel peak appears near the wavelength where $k$ reaches its maximum for crystalline ice, which would normally be associated with strong absorption. However, by taking the partial derivatives of the Rayleigh-regime scattering factor, $\left|\frac{m^2-1}{m^2+2}\right|^2$, in equation~\eqref{kappa_sca_Rayleigh} with respect to $n$ and $k$, and using the fact that $n \sim k$ near 3.1~\mum, we find that this factor is a decreasing function of $n$ and an increasing function of $k$ in the vicinity of 3.1~\mum. Therefore, in the Rayleigh regime, $\kappa_{\rm sca}$ exhibits a peak around 3.1~\mum, where $n$ has a local minimum and $k$ has a local maximum. This shows that, near the O--H stretching resonance, the increase in $k$ does not simply enhance absorption but, together with the accompanying variation in $n$, also strengthens the induced polarization within the particle and thereby enhances the scattering response. The ordered molecular structure of crystalline ice makes this polarization response more pronounced and thereby gives rise to the Fresnel peak.

A similar argument can also be made in the geometric-optics regime. For simplicity, considering normal incidence, the reflectance in equation~\eqref{Reflectance} is written as $\frac{(n-1)^2+k^2}{(n+1)^2+k^2}$. This is an increasing function of $k$ and, when $n^2-1<k^2$, a decreasing function of $n$, so that $\kappa_{\rm sca}$ also peaks at $\sim 3.1$~\mum\ in the geometric-optics regime. Even for large particles, the crystalline structure enhances the polarization response and increases the reflectance at the particle surface. As a result, scattered light from crystalline ice exhibits a Fresnel peak irrespective of particle size.

To validate the above estimations, Figure~\ref{fig:optical_property_Mie} also shows $\kappa_{\rm abs}$, $\kappa_{\rm sca}$, $\kappa_{\rm ext}$, and $\omega$ computed using Mie calculations with \texttt{OpTool} \citep{2021ascl.soft04010D}. Although the Mie results are compared with the analytic expressions described above, for large grains we apply analytic expressions based on anomalous diffraction theory outside the 2.9--3.3~\mum\ (see Appendix~\ref{append:ADtheory}).
The results agree with the trends predicted by equations ~\eqref{kappa_abs_Rayleigh}--\eqref{kappa_sca_GO}, supporting the conclusion that the Fresnel feature provides a useful diagnostic of crystalline ice across a broad range of grain sizes. \footnote{In the Mie calculations, we adopt a log-normal grain-size distribution with mean grain size of 0.1 or 100 \mum\ and a dispersion of 0.03, truncated at 0.08 and 120 \mum. This choice suppresses oscillatory behavior in the optical properties that becomes prominent for grains larger than 10 \mum\ due to Mie interference fringes (see $\kappa_{\rm ext}$ around 2.6 \mum). We verified that the small discrepancies between the analytic expressions and the Mie results can be reduced by narrowing the log-normal distribution; however, this comes with the trade-off that the interference-induced oscillations become more prominent.\label{fot:interference}}
In the following sections, we perform synthetic imaging of planet-forming disks using dust optical properties derived from the Mie calculations. The resulting scattered-light spectra reflect these optical properties as well as the disk dust density distribution and scattering geometry, and—in the case of protoplanetary disks—additional effects such as multiple scattering.

\section{Simulation Methods}
\label{sec:simulation_method}
\subsection{Optical properties of dust grains}
For disk imaging, we specify the dust size and composition and compute the grains' scattering and absorption opacities, $\kappa_{\rm sca}$ and $\kappa_{\rm abs}$, and the elements $S_{ij}$ of the scattering matrix $S$. We use \texttt{OpTool} \citep{2021ascl.soft04010D} to calculate these properties, adopting standard Mie theory for homogeneous spheres. As dust constituents, we consider two components: rocks (mixture of silicate and organics) and water ice. The calculations require the complex refractive indices of these materials; for rocks we use \texttt{astrodust} from \citet{2021ApJ...909...94D,2023ApJ...948...55H}, adopting $b/a = 1.4$, $P = 0.2$, and $f_{\rm Fe} = 0$ for the spheroidal shape, porosity, and metallic Fe inclusion fraction, respectively, and for water ice we use crystalline and amorphous H$_2$O from \citet{2009ApJ...701.1347M}. In this study, we vary the grain size, the rock-to-ice mass ratio, and the degree of ice crystallinity to quantify how these parameters influence the near- to mid-infrared scattered-light spectra of planet-forming disks and the associated H$_2$O features. The specific parameter choices are as follows.

\begin{description}[style=nextline,font=\bfseries,leftmargin=!]
    \item[Dust size $a$] 
    Grain size affects the near-infrared spectral slope and, consequently, the depth and shape of the H$_2$O features \citep[e.g., ][]{2008PASJ...60..557I, 2021ApJ...921..173T}, including the Fresnel peak. To examine the effect of grain size on the spectra, we first adopt narrow log-normal size distributions with mean sizes of 0.1, 1, 10, and 100~$\mu$m and a dispersion of 0.3, where the grain size $a$ refers to the mean of the distribution. We avoid purely single-sized grains primarily to suppress oscillatory structures in the scattered-light spectra caused by Mie interference, which become prominent for grains with $a \gtrsim 10~\mu$m. We also examine a case with a size distribution: a power-law grain-size distribution over 0.01--100 $\rm{\mu m}$ with $n(a)da \propto a^{-p}da$, where $n(a)da$ is the number density of particles with sizes in $[a,a+da]$ and $p$ is the power-law index of the grain size distribution. We consider $p = 3.5$ and $p = 2.5$.

    \item[Rock-to-ice mass ratio $f_{\rm rock}$]  
    We adopt $f_{\rm rock}=0.3, 1, 3$ to assess the impact on the spectra.

    \item[Crystallinity $c$] 
    We consider $c=0$ (pure amorphous ice), 0.25, 0.5, 0.75, and 1.0 (pure crystalline ice).
\end{description}
In the debris-disk simulations, we consider multiple combinations of the rock-to-ice mass ratio $f_{\rm rock}$ and disk inclination $i$. For each $(f_{\rm rock}, i)$ pair, we run a parameter study by varying the grain size and crystallinity. In total, this yields 78 debris-disk models, and we define the fiducial model as $f_{\rm rock}=1$ and $i=0^\circ$. Because the PPD simulations are more computationally expensive than the debris-disk runs, we restrict our PPD simulations to six cases: we fix $f_{\rm rock}=1$ and adopt a power-law grain-size distribution over 0.1--100~$\mu$m with index $p=3.5$, while varying only the inclination and crystallinity.
The combinations of the above parameters used for the dust grains in our disk-imaging simulations are summarized in Table \ref{tab:parameter_set}.

\subsection{Disk imaging}
\subsubsection{Debris Disk}
\label{method:subsec:debris_disk}
We compute scattered-light images of debris disks using the public code \texttt{DDiT} \citep{2020A&A...640A..12O}. For disk imaging, \texttt{DDiT} first defines a three-dimensional spheroidal computational domain that encloses the bulk of the disk dust. Spatial sampling points are then placed within this domain, and the local dust spatial density $\rho$ and the scattering angle $\theta$ are evaluated at each point. The scattering phase function is obtained by linear interpolation of the prescribed scattering-matrix element $S_{11}$ on a grid of scattering angles. At each sampling point, the contribution to the specific intensity estimated as $\propto \rho\,S_{11}(\theta)$. The final image is obtained by integrating these contributions along the line of sight. The resulting spectrum corresponds to the surface brightness divided by the stellar flux.
We set the number of pixels to 300 and the number of sampling points along the line of sight to 150.

In cylindrical coordinates (r,z), the dust spatial-density distribution $\rho(r,z)$ is given by:
\begin{equation}
    \rho(r,z) \propto \left[\left(\frac{r}{r_0} \right)^{-2\alpha_{\rm in}} + \left(\frac{r}{r_0} \right)^{-2\alpha_{\rm out}} \right] \exp\left[\left(-\frac{z}{z_{\rm disk}}\right)^2\right],
\end{equation}
where, $r_0$ is the reference radius, $\alpha_{\rm in}$ and $\alpha_{\rm out}$ are the radial density slopes on the inner and outer sides of the disk, respectively, and $z_{\rm disk}$ is the disk height. For simplicity, we adopt a disk with reference radius $r_0 = 1''$ without eccentricity. We set $\alpha_{\rm in} = 25$ and $\alpha_{\rm out} = -2.5$, corresponding to a profile with a sharp density drop toward the inner edge. We note that in \texttt{DDiT} the scattered-light intensity is evaluated from $\rho$ and $S_{11}(\theta)$, so no specific stellar properties need to be assumed.
To clarify the impact of dust optical properties on the scattered-light spectra, we first assume a completely face-on disk (inclination $i=0^{\circ}$) and perform imaging using the dust optical properties listed in Table \ref{tab:parameter_set}. We then investigate the effect of disk inclination on the spectra by imaging for $i=45^{\circ}$ and $i=90^{\circ}$.

\begin{table*}[t]
\centering
\setlength{\tabcolsep}{4pt}
\caption{Parameter set for models.}
\label{tab:parameter_set}
\begin{tabularx}{\textwidth}{@{}l l l l >{\raggedright\arraybackslash}X@{}}
\toprule
\textbf{Disk type} &
\textbf{Geometry} &
\textbf{Grain size $a$} &
\makecell[l]{\textbf{Rock-to-ice} \\ \textbf{mass ratio $f_{\rm rock}$}} &
\textbf{Crystallinity $c$} \\
\midrule
\multirow{7}{*}{Debris disk}
  & \multirow{5}{*}{Face-on ($i=0^\circ$)}
  & log-normal: $0.1~\mu\mathrm{m}$ 
  & \multirow{4}{*}{$0.3$, $1$, $3$}
  & \multirow{4}{=}{for $f_{\rm rock}=1$: $c=\{0,\,0.25,\,0.5,\,0.75,\,1.0\}$; otherwise: $c=\{0,\,0.5,\,1.0\}$} \\
  & & log-normal: $1~\mu\mathrm{m}$   & & \\
  & & log-normal: $10~\mu\mathrm{m}$  & & \\
  & & log-normal: $100~\mu\mathrm{m}$ & & \\
\cmidrule(lr){3-5}
  &  & $0.1$--$100~\mu\mathrm{m}$ / $p=\{3.5,\,2.5\}$ & $1$
  & $c=\{0,\,0.25,\,0.5,\,0.75,\,1.0\}$ \\
\cmidrule(lr){2-5}
  & Inclined ($i=45^\circ,\,90^\circ$)
  & log-normal sizes as above & $1$ & $c=\{0,\,0.5,\,1.0\}$ \\
\midrule
\multirow{2}{*}{Protoplanetary disk}
  & Face-on ($i=5^\circ$)
  & \multirow{2}{*}{$0.1$--$100~\mu\mathrm{m}$ / $p=3.5$}
  & \multirow{2}{*}{$1$ (if $T < 150$ K)}
  & \multirow{2}{*}{$c=\{0,\,0.5,\,1.0\}$} \\
  & Edge-on ($i=85^\circ$)
  & & & \\
\bottomrule
\end{tabularx}
\end{table*}

\subsubsection{Protoplanetary disk}
Although it is still unclear whether the Fresnel peak observed in debris disks can also be reported in protoplanetary disks, it is important to clarify the properties of crystalline water-ice features in protoplanetary-disk spectra, particularly given the anticipated increase in high-resolution observations with JWST and similar facilities. We therefore compute the disk temperature structure and perform imaging with the three-dimensional Monte Carlo radiative-transfer code \texttt{RADMC-3D} \citep{2012ascl.soft02015D}.

The dust surface density of the PPD is given by
\begin{equation}
    \Sigma(r) = \delta\,\Sigma_0\left(\frac{r}{r_0} \right)^{-\beta}
    \exp\left[-\left(\frac{r}{r_t}\right)^{2-\beta}\right],
\end{equation}
where $r_0=100$~au is the reference radius, $r_t=300$~au is the cutoff radius, and $\beta=0.7$ sets the radial dependence of the surface density. The factor $\delta$ is introduced to mimic a transition-disk structure, as observed in many disks in recent years, and is defined as
\begin{equation}
\delta =
\begin{cases}
0.01, & r < 10~\mathrm{au},\\
1, & r \ge 10~\mathrm{au}.
\end{cases}
\end{equation}
The reference surface density $\Sigma_0$ is chosen such that the total dust mass equals $2\times10^{-5}M_{\odot}$. The dust spatial density is
\begin{equation}
    \rho(r,z) = \frac{\Sigma(r)}{\sqrt{2\pi}H(r)} \exp\left(-\frac{z^2}{2H^2(r)}\right),
\end{equation}
where $H(r)$ is the disk scale height, given by
\begin{equation}
    H(r) = H_0\left(\frac{r}{r_0}\right)^\psi,
\end{equation}
where $H_0 =7.5$ au is the reference value of $H$ at $r_0$ and $\psi = 1.2$  sets the radial dependence of the scale height.
We assume the axisymmetric disk and use a 3D spherical coordinate grid with cells of $256\times128\times128$.
Our choice of the power-law indices that define the disk density structure follows the profile proposed for Tau 042021, as observed with JWST, excluding its extended-atmosphere component \citep{2024AJ....167...77D}.
We assume a disk around a T~Tauri star, adopting a stellar radius of $2\,R_\odot$, a mass of $0.5\,M_\odot$, and an effective temperature of $4000$~K.

Unlike debris disks, protoplanetary disks typically exhibit temperature variations spanning roughly three orders of magnitude from the inner to the outer regions. We first assign to the entire disk the optical properties of \texttt{astrodust} with a grain-size distribution of 0.1--100 \mum, calculated with \texttt{OpTool}, and compute the dust temperature with \texttt{RADMC-3D}. Based on this temperature structure, we divide the disk into several temperature regions and reassign optical properties for imaging as follows. In regions with $T>150$ K, we assume that ice is absent and apply the optical properties of \texttt{astrodust}. For the interval $20 ~\rm{K}<T<150$ K, we assume dust with an rock-to-ice mass ratio of unity. We then bin this region in steps of 10 K and assign optical properties computed using the complex refractive indices of \texttt{astrodust} and H$_2$O (the latter from Mastrapa et al. 2009) at the corresponding temperatures. To avoid numerical artifacts in the scattered-light images caused by the combination of an extremely forward-peaked phase function from large grains with $x \gg 1$ and finite spatial resolution, we cap the scattering-matrix values at small scattering angles in the \texttt{OpTool} calculations by setting them equal to the value at a scattering angle of $5^\circ$. We adopt $10^7$ photon packets for both the temperature calculation and the imaging.

\subsection{Analysis of scattered light spectra}
The most prominent diagnostic of the H$_2$O ice phase is the Fresnel peak appearing near 3.1~$\rm{\mu m}$ (Figure~\ref{refractive_index}). In addition, crystalline ice exhibits a feature near 3.25~$\rm{\mu m}$ (hereafter the "secondary peak"), which is particularly pronounced in backscattering regimes. One of our goals is to quantify how ice crystallinity, the ice/rock mass ratio, and grain size affect these features. Strictly speaking, the term "Fresnel peak" can refer not only to the sharp local maximum produced by crystalline ice, but also to the broad maximum around 3.0~$\rm{\mu m}$ associated with amorphous ice, as seen in $\kappa_{\rm sca}$ in Figure~\ref{fig:optical_property_Mie} \citep{2009ApJ...701.1347M}. In this work, however, we focus on the sharp crystalline-ice features that can be clearly identified in observed spectra.

As an observable, model-independent metric that can be directly compared with data, we introduce 
\begin{equation} 
\label{eq:feature_height} 
\Delta_i \equiv \ln \left( \frac{I_i}{I_{\rm cont}} \right), 
\end{equation} 
where the subscript $i$ denotes either the Fresnel peak or the secondary peak (i.e., $i \in \{\mathrm{Fresnel},\,\mathrm{secondary}\}$), $I_i$ is the scattered-light intensity at the Fresnel peak ($\sim$3.1~$\rm{\mu m}$) or at the secondary peak ($\sim$3.25~$\rm{\mu m}$), and $I_{\rm cont}$ is the intensity of a pseudo-continuum obtained by masking the vicinity of each peak and interpolating the remaining spectrum with a polynomial function. We restrict the masking windows used for the continuum interpolation to the vicinities of 3.1 and 3.25~$\rm{\mu m}$ so that $\Delta_i$ is close to zero for fully amorphous ice. We refer the reader to Appendix~\ref{append:continuum_const} for details of the pseudo-continuum construction and illustrative examples. In practice, we identify the spectral grid point at which the difference between the original spectrum and the constructed continuum is maximized, and evaluate $\Delta_i$ at that grid point. Note that even at zero crystallinity, the interpolated continuum does not perfectly match the model spectrum, and therefore $\Delta_i$ is not exactly zero.

\section{Ice feature in debris disks}
\label{sec:debri_res}
\subsection{Scattered light images and spectra}
\label{subsec:debri_imaging}
\subsubsection{Face-on debris disks}
\label{subsubsec:debri:face-on}

%%%%%%%%%%%%%%%%%%%%%%%%%%%%%%%%%%%
\begin{figure*}[p]
\includegraphics[width=0.80\hsize]{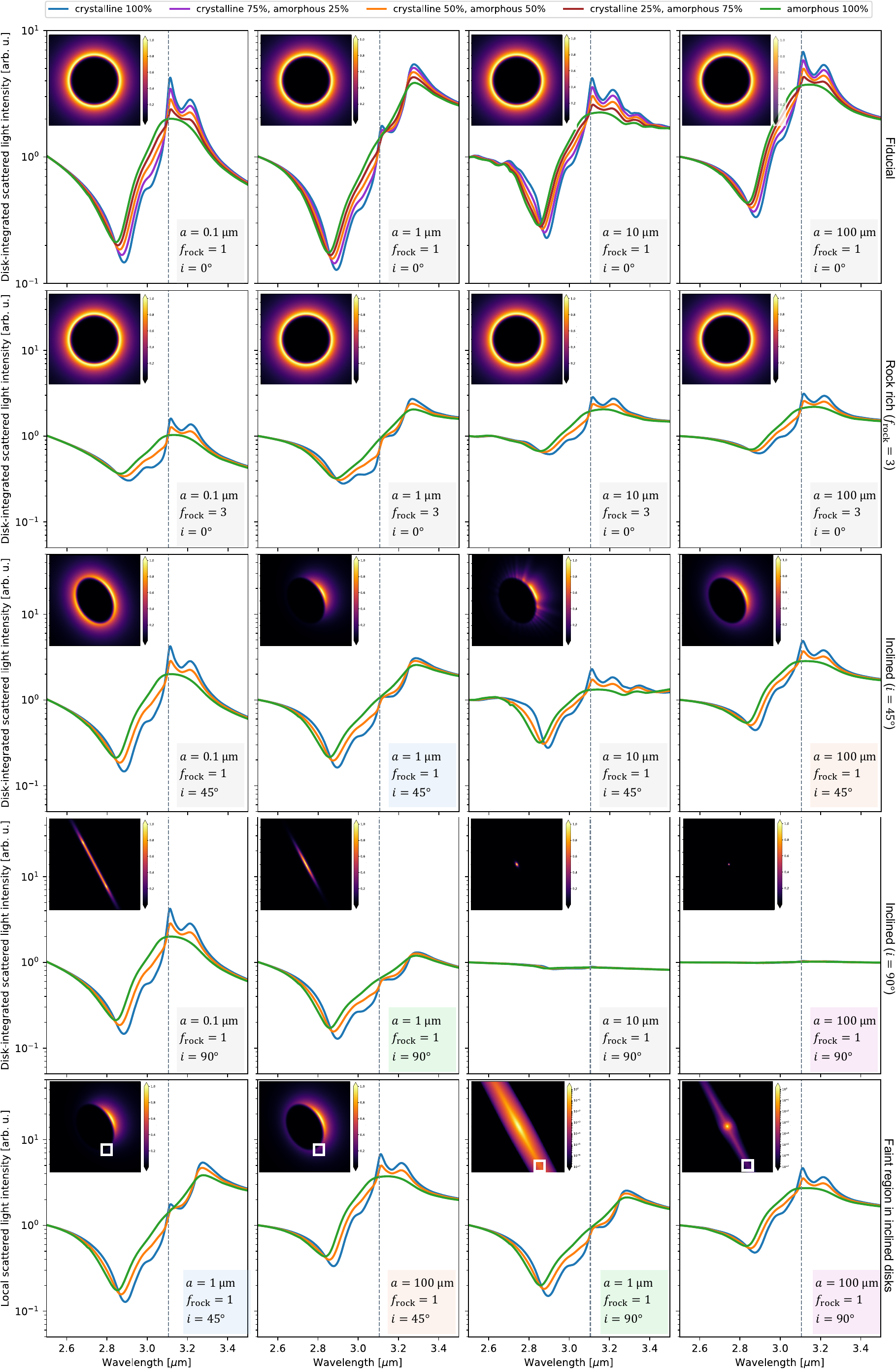}
\centering
\caption{Integrated scattered-light spectra of debris disks for different grain sizes, rock-to-ice mass ratios, and disk inclinations. The spectra are normalized to the value at 2.5~\mum. From top to bottom, the rows show the fiducial case, the rock-rich case ($f_{\rm rock}=3$), the inclined-disk case ($i=45^\circ$), the edge-on case ($i=90^\circ$), and the spectra extracted from the $90^\circ$-scattering regions in the inclined and edge-on cases. The corresponding scattered-light images are shown in the inset. Each image is normalized by its maximum brightness, except for the two edge-on panels in the bottom row, which are shown in logarithmic scale without normalization. {Alt text: Scattered-light images and spectra of debris disks.}}
\label{fig:debri_spectra}
\end{figure*}
%%%%%%%%%%%%%%%%%%%%%%%%%%%%%%%%%%

First, we present synthetic imaging results for face-on debris disks and overview how the scattered-light spectrum varies with grain size and ice crystallinity.
The top row of Figure~\ref{fig:debri_spectra} shows synthetic images of the debris disk in the fiducial case, normalized by the maximum surface brightness in each image. The absolute peak brightness is highest for the  $a=0.1$ \mum\ case and decreases approximately inversely with grain size, while all images have similar axisymmetric structure. As predicted by the analytic considerations in Section~\ref{sec:fresnel_feature}, the integrated scattered-light spectra are broadly similar for both the smallest grains ($a=0.1$ \mum) and the large-grain cases ($a\gtrsim 10$ \mum), and crystalline ice produces a 3.1~\mum\ Fresnel peak and a 3.25~\mum\ secondary peak (Figure~\ref{fig:debri_spectra}, top row). Both peaks become more pronounced with increasing crystallinity. Only for $a=1$ \mum, for which $x\sim1$, does a positive spectral slope extend from the 2.9~\mum\ O--H stretching absorption out to $\sim$3.3 \mum, with the Fresnel feature superposed on this rising continuum. As a result, a high crystallinity ($c\gtrsim0.75$) is required for the Fresnel feature to emerge as a distinct peak; at lower crystallinity it appears instead as a step-like kink. We also verified that grains following a power-law size distribution yield spectra with a pronounced Fresnel peak resembling the $a=10$ \mum\ log-normal case, for both $p=3.5$ and $p=2.5$.

For more rock-rich grains, a bowl-like absorption feature emerges between 2.8 and 3.1~\mum, similar to that observed with JWST for TNOs \citep{2025NatAs...9..245L} (Figure~\ref{fig:debri_spectra}, second row). This feature becomes clearer at higher crystallinity. As in the fiducial case, all sizes except $a=1$ \mum\ show a Fresnel peak, although with reduced amplitude. For $a=1$ \mum\, even fully crystalline ice does not produce a distinct peak; instead, the Fresnel feature remains a kink. Conversely, for more ice-rich compositions than the fiducial case, the overall spectral morphology is nearly unchanged, but the Fresnel peak becomes higher.

\subsubsection{Inclined debris disks}
\label{subsubsec:debri:inclination}

In Section \ref{subsubsec:debri:face-on}, we simulate scattered-light spectra for a face-on debris disk. In practice, however, observed disks span a range of inclinations. Here we investigate how disk inclination affects the scattered-light spectra and, in particular, the crystalline-ice features.
In contrast to the face-on case, inclined debris disk images exhibit grain-size-dependent morphologies (Figure \ref{fig:debri_spectra}, third and fourth rows). At $i=45^\circ$, only the very small-grain case ($a < 1$ \mum) remains nearly axisymmetric, whereas for $a \gtrsim 1$ \mum\ the image clearly separates into a bright forward-scattering side and a faint backward-scattering side. The $a=10$ \mum\ image shows a dark spike-like structure, which arises from Mie interference (see footnote \ref{fot:interference}). Although the 2.8--3.1 \mum\ bowl-shaped attenuation characteristic of the rock-rich, face-on case is not present, the Fresnel feature exhibits trends similar to those in the rock-rich, face-on spectra (Figure \ref{fig:debri_spectra}, third row). Specifically, depending on the crystallinity, the $a=1$ \mum\ case exhibits a kink-like Fresnel feature, whereas the other sizes produce a Fresnel peak (see Section \ref{subsec:debri:features_height} for a quantitative evaluation of the feature heights).

In the edge-on case ($i=90^\circ$), small grains ($a < 1$ \mum) produce an elongated, bar-like morphology, reminiscent of that observed in systems such as $\beta$~Pictoris \citep[e.g., ][]{1984Sci...226.1421S,2000ApJ...539..435H, 2006AJ....131.3109G}. For larger grains, the scattered light becomes concentrated toward the central region due to forward scattering, yielding a nearly point-source-like image. The integrated scattered-light spectra for $a < 1$ \mum\ are similar to those in the face-on and $i=45^\circ$ cases, whereas for larger grains the spectra become nearly flat, making the Fresnel feature difficult to identify (Figure \ref{fig:debri_spectra}, fourth row).

In inclined disks, the disk integrated flux is dominated by the contribution from bright forward-scattered light, making the Fresnel feature less distinct than in the face-on case, where the observed scattered light mainly arises from regions with scattering angles around $90^\circ$. This is consistent with the results of \citet{2024MNRAS.533.2801K}, who showed that the Fresnel feature is more strongly enhanced in backward scattering. In actual observations, however, the contribution from the bright central source may be suppressed by a coronagraph, allowing detection of the fainter scattered light originating from regions with scattering angles of approximately $90^\circ$. The bottom row of Figure~\ref{fig:debri_spectra} shows spectra selectively extracted from regions with scattering angles of $\sim 90^\circ$. The selected regions are enclosed by white boxes, and for clarity the images of the edge-on debris disk are shown again on a logarithmic scale without normalization. By isolating the scattered light at scattering angles around $90^\circ$, the Fresnel feature becomes more pronounced than in the corresponding disk integrated flux. In particular, for the case of an inclination of $45^\circ$ and $a = 1$ \mum, the disk integrated flux exhibits only a kink-like Fresnel feature even when the crystallinity is $c = 1$, whereas the flux extracted from the $\sim90^\circ$ scattering region shows a distinct Fresnel peak. Likewise, whereas the disk integrated flux of the edge-on debris disk becomes flat for larger grains because it is dominated by centrally concentrated, featureless forward-scattered light, the $\sim 90^\circ$ scattered light leaking out from the central region yields a spectrum that contains a Fresnel feature similar to that in the face-on case.

\subsection{Evaluation of crystalline H$_2$O features}
\label{subsec:debri:features_height}
%%%%%%%%%%%%%%%%%%%%%%%%%%%%%%%%%%%
\begin{figure*}[t]
\centering
\includegraphics[width=0.8\hsize]{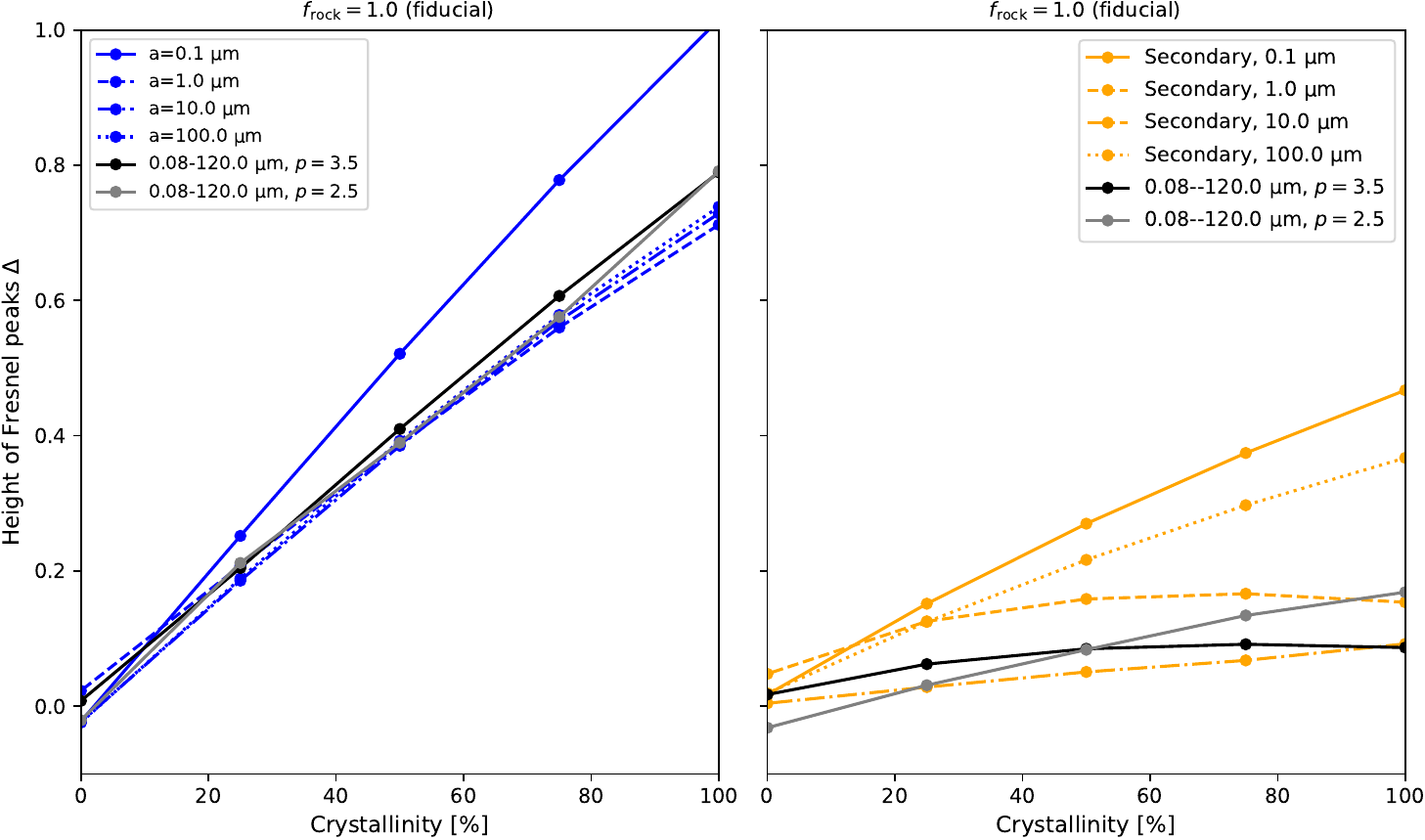}
\caption{Heights of the Fresnel and secondary peaks in the integrated scattered-light spectra of the debris disks shown in the fiducial case.
Left: Height of the Fresnel peak, where the masking wavelength range used for continuum construction is fixed at 3.05--3.2 \mum ~for all spectra.
Right: Height of the secondary peak, where the masking wavelength range for continuum fitting is adjusted individually for each spectrum within 3.1--3.3 \mum ~so that the spline interpolation works smoothly. {Alt text: Two line graphs showing the height of crystalline ice features.}}
\label{fig:feature_height_fid}
\end{figure*}
%%%%%%%%%%%%%%%%%%%%%%%%%%%%%%%%%%

%%%%%%%%%%%%%%%%%%%%%%%%%%%%%%%%%%%
\begin{figure*}[t]
\centering
\includegraphics[width=0.8\hsize]{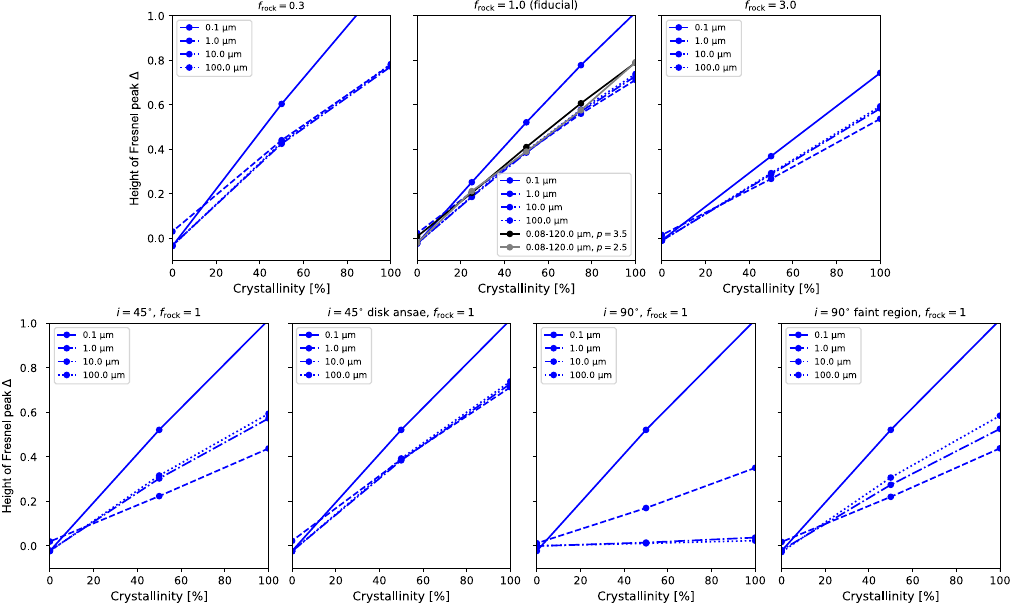}
\caption{Heights of the Fresnel peaks in the scattered-light spectra of debris disks composed of different rock-to-ice ratios and inclinations. The top panels show the values measured from the disk integrated flux of face-on debris disks for rock-to-ice mass ratios of 0.3, 1 (fiducial), and 3. The bottom panels show the values measured from the disk integrated flux and from the faint $90^\circ$ scattered-light spectra of inclined and edge-on debris disks. {Alt text: Line graphs showing the height of Fresnel feature in various debris disk models.}}
\label{fig:Fresnel_params}
\end{figure*}
%%%%%%%%%%%%%%%%%%%%%%%%%%%%%%%%%%

%%%%%%%%%%%%%%%%%%%%%%%%%%%%%%%%%%%
\begin{figure}[t]
\includegraphics[width=0.85\hsize]{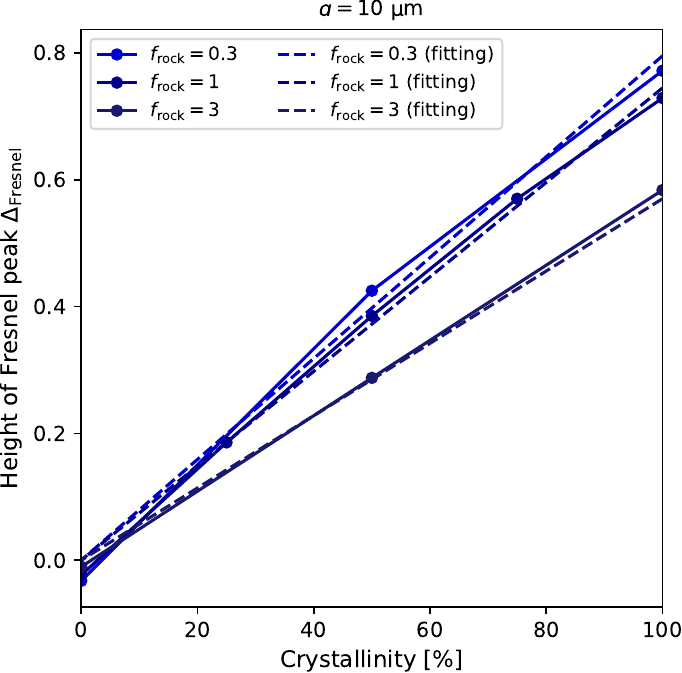}
\caption{Comparison between $\Delta_{\rm Fresnel}$ for 10 \mum ~grains with different rock-to-ice mass ratios and the fitting function \eqref{eq:fitting_delta_fresnel}. {Alt text: A line graph showing the performance of equation \eqref{eq:fitting_delta_fresnel}.}}
\label{fig:comparison_fit_and_model}
\end{figure}
%%%%%%%%%%%%%%%%%%%%%%%%%%%%%%%%%%

In Section \ref{subsec:debri_imaging}, we found that for face-on debris disks and the $\sim 90^\circ$-scattering regions of inclined debris disks, the scattered-light spectral morphology is broadly similar across grain sizes, and that the Fresnel and secondary peak heights exhibit clear, crystallinity-dependent variations across all cases spanning different grain sizes and rock-to-ice mass ratios. Motivated by the goal of establishing an observationally comparable indicator of ice crystallinity for debris-disk scattered light at scattering angles of $\sim 90^{\circ}$, we quantify these peaks using the metric defined in equation \eqref{eq:feature_height}. Figure \ref{fig:feature_height_fid} shows the Fresnel- and secondary-peak heights in the fiducial case as functions of crystallinity. First, the Fresnel-peak height, $\Delta_{\rm Fresnel}$, increases approximately linearly with crystallinity, largely independent of grain size and of whether the grains have a size distribution. Although the single-size 0.1 \mum ~case yields a larger $\Delta_{\rm Fresnel}$ than the others, it is striking that for the remaining sizes, including those with a size distribution, $\Delta_{\rm Fresnel}$ v.s. crystallinity fall nearly on the same trend. Moreover, in constructing the pseudo-continuum, the Fresnel peak can be handled uniformly by masking 3.05-3.20 \mum ~and applying a cubic-spline interpolation, indicating that the Fresnel peak provides a well-behaved quantitative diagnostic.

By contrast, although the secondary-peak height, $\Delta_{\rm secondary}$, is generally an increasing function of crystallinity, it is not linear and exhibits larger size-dependent variation than $\Delta_{\rm Fresnel}$. As with the Fresnel peak, the secondary peak is most prominent for 0.1 \mum ~grains, whereas its dependence on crystallinity is weak for 10 \mum ~grains. In addition, the secondary peak’s central wavelength shifts slightly with crystallinity and the feature is broader than the Fresnel peak, so constructing a smooth continuum requires fine-tuning of the masking wavelengths. For this reason, the secondary peak is a less stable quantitative indicator of crystallinity than the Fresnel peak.

The upper row of Figure \ref{fig:Fresnel_params} shows $\Delta_{\rm Fresnel}$ as a function of crystallinity for different rock-to-ice mass ratios ($f_{\rm rock}=0.3$--3). When the grains are mixtures of rock and ice, regardless of which component is dominant, $\Delta_{\rm Fresnel}$ increases approximately linearly with crystallinity. The size dependence is similar to the fiducial case: only the $a=0.1$ \mum ~grains exhibit a large increase in $\Delta_{\rm Fresnel}$ with crystallinity, whereas the other sizes fall nearly on the same trend. Additionally, as the rock-to-ice mass ratio increases, $\Delta_{\rm Fresnel}$ becomes smaller at any given grain size and crystallinity, and its dependence on crystallinity (i.e., the slope) also decreases slightly.

The $\Delta_{\rm Fresnel}$ values measured from the disk integrated flux of inclined disks show a linear relationship with crystallinity quantitatively similar to that in the fiducial case only when the disk is composed of small grains with $a = 0.1$ \mum~(Figure \ref{fig:Fresnel_params}, lower row). For larger grain sizes, $\Delta_{\rm Fresnel}$ becomes smaller than in the fiducial case even at the same crystallinity. Moreover, in edge-on disks composed of grains with $a > 10~\mu$m, for which the disk integrated flux become nearly flat, $\Delta_{\rm Fresnel}$ is close to zero for any crystallinity. By contrast, it is noteworthy that even in inclined disks, when scattered light at scattering angles around $90^\circ$ is selectively extracted, a linear $\Delta$--$c$ relationship similar to that in the fiducial case is recovered. For the scattered-light spectra extracted from the disk ansae of the debris disk at $i=45^\circ$, the $\Delta_{\rm Fresnel}$--$c$ relationship is quantitatively consistent with that in the fiducial case. Even for $i = 90^\circ$, the faint-region scattered-light spectra still exhibit a linear $\Delta$--$c$ relationship independent of grain size, although the slope is slightly smaller than in the fiducial case, because the line-of-sight integration inevitably includes contributions from scattering at angles other than $90^\circ$.

Because scattered light at scattering angles around $90^\circ$ is relatively accessible observationally, it is worthwhile to establish an empirical relation that allows the crystallinity to be inferred from such scattered-light spectra. Building on the dependences of $\Delta_{\rm Fresnel}$ on crystallinity $c$ and on the rock-to-ice mass ratio $f_{\rm rock}$ in face-on disks, we therefore seek to express $\Delta_{\rm Fresnel}$ as a function of these parameters. Empirically, $\Delta_{\rm Fresnel}$ varies approximately linearly with $c$. The linearity of the $\Delta_{\rm Fresnel}$--$c$ relation is preserved across different $f_{\rm rock}$, while the absolute value of $\Delta_{\rm Fresnel}$ at a given $c$ decreases as $f_{\rm rock}$ increases. Moreover, for grain sizes $a\gtrsim 1$~\mum, the $\Delta_{\rm Fresnel}$--$c$ relation is nearly independent of grain size, i.e., the peak height at a given crystallinity varies little across sizes. Physically, the relation should satisfy $\Delta_{\rm Fresnel}=0$ at $c=0$, and $\Delta_{\rm Fresnel}\rightarrow 0$ as $f_{\rm rock}\rightarrow \infty$. Using all $\Delta_{\rm Fresnel}$--$c$ data for grains with $a\gtrsim1$ \mum, we therefore fit the parameters $\Delta_{\rm Fresnel, bf}$, $f_{\rm rock,bf}$, and $\gamma$ in the following relation by least squares:
\begin{equation}
\label{eq:fitting_delta_fresnel}
\Delta_{\rm Fresnel} = c\Delta_{\rm Fresnel, bf}\exp{\left[-\left(\frac{f_{\rm rock}}{f_{\rm rock,bf}}\right)^\gamma\right]}.
\end{equation}
From our dataset we obtain $[\Delta_{\rm Fresnel, bf}, f_{\rm rock,bf}, \gamma]=[0.80, 6.8, 1.3]$. Figure \ref{fig:comparison_fit_and_model} compares $\Delta_{\rm Fresnel}$ derived from integrated scattered-light spectra of grains with $a=10$ \mum ~at different $f_{\rm rock}$ with the fitting function \eqref{eq:fitting_delta_fresnel}, showing that the fit reproduces the dependences of $\Delta_{\rm Fresnel}$ on both crystallinity and rock-to-ice mass ratio. Although this empirical relation was established from simulations of face-on debris disks, it is expected to be applicable also to estimating the crystallinity of inclined disks when scattered light at scattering angles around $90^\circ$ is selectively extracted.

Note that in our simulations we adopt laboratory-measured complex refractive indices of crystalline and amorphous water ice at $T=50$~K. The temperature dependence of the Fresnel feature is discussed in Appendix~\ref{append:temperature_effect}. Our fitting function is applicable to ice in cold environments; relevant real systems include debris disks at $r \gtrsim 10$~au around solar analogues, as well as long-period comets and icy small bodies.

%%%%%%%%%%%%%%%%%%%%%%%%%%%%%%%%%%%
\begin{figure*}[t]
\includegraphics[width=0.8\hsize]{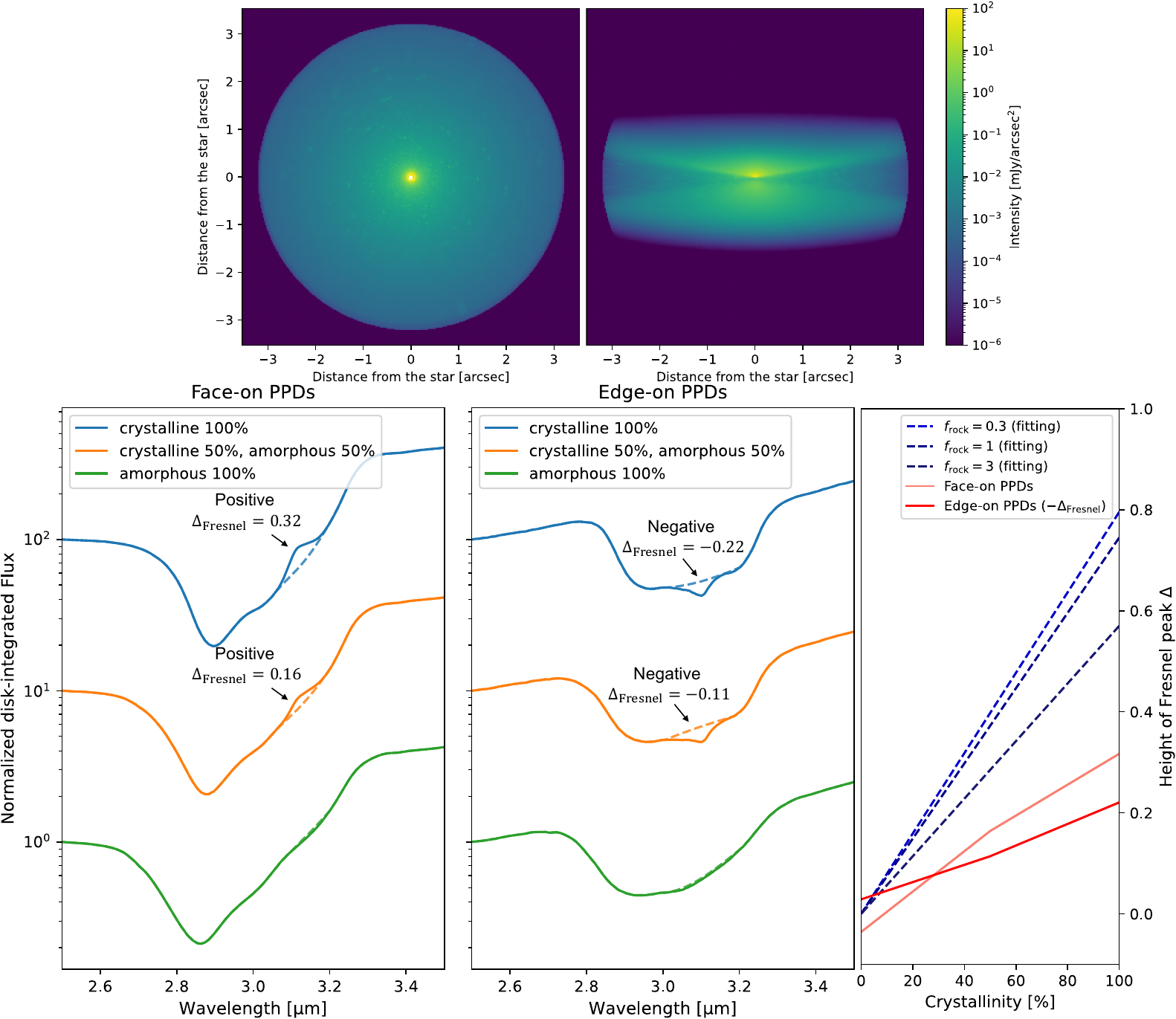}
\centering
\caption{Protoplanetary disk images (top row) and scattered-/transmitted-light spectra (bottom-left and bottom-middle) obtained from the radiative-transfer simulations, together with the crystallinity dependence of their Fresnel-feature heights (bottom-right). For the edge-on case, the Fresnel-feature height is measured in absolute value of feature depth. {Alt text: Theoretical images, spectra, and Fresnel feature height of the protoplanetary disks.}}
\label{fig:PPDs_spectra}
\end{figure*}
%%%%%%%%%%%%%%%%%%%%%%%%%%%%%%%%%%

\section{Ice feature in protoplanetary disk}
\label{sec:ppds_res}
Whereas highly crystalline ice in cold debris disks produces a prominent Fresnel peak at 3.1~\mum, it remains possible that a similar feature in protoplanetary disks could be detected with high-sensitivity facilities such as JWST. In this section, we use radiative-transfer calculations with \texttt{RADMC-3D} to demonstrate how the scattered-/transmitted- light spectra of protoplanetary disks vary with ice crystallinity and disk geometry.

\subsection{Images and spectra}
\label{subsec:ppd_spectra}
The lower left panel of Figure \ref{fig:PPDs_spectra} shows the disk integrated flux of a face-on protoplanetary disk, with the corresponding images at 2.0 \mum~ shown in the upper left panel. To mimic the effect of a coronagraph in actual observations, which suppresses stellar light, we set the flux in pixels within $\sim 0.1^{\prime\prime}$ of the central star to zero. For a face-on protoplanetary disk, the disk integrated flux exhibits an H$_2$O feature around 3 \mum, similar to that seen in debris disks. This feature has a V-shaped profile centered near 2.9 \mum~ and extends to $\sim$3.4 \mum; when the ice is entirely amorphous, the flux rises smoothly from 2.9 to 3.4 \mum. As the crystallinity increases, a kink-like structure seen in rock-rich debris disks appears near 3.1 \mum~ due to Fresnel feature from crystalline ice. This behavior more closely follows the scattering albedo, $\omega$, rather than the scattering opacity, $\kappa_{\rm sca}$. This is consistent with the analytic argument by \citet{2008PASJ...60..557I}, who showed that the scattered-light spectrum in optically thick disks scales approximately with the scattering albedo under the assumption of isotropic scattering. Hereafter, we refer to the Fresnel feature that scales with the scattering albedo as the \textit{albedo feature}. In addition, PPD spectra include scattered light from the inner disk, where grains are ice-free. This dilutes the contribution from icy dust and tends to weaken water-ice features. 

Compared to the face-on case, the edge-on PPD spectrum is characterized by a broad, bowl-shaped absorption extending from 2.9 to 3.4~\mum\ (Figure \ref{fig:PPDs_spectra}, lower middle). Because an edge-on geometry predominantly probes transmitted light through the disk, the spectrum more closely resembles the inverse of the extinction opacity. A key signature of crystalline ice is the addition of a small  absorption near 3.1~\mum\ superposed on the broad bowl-shaped feature, which becomes more pronounced as the crystallinity increases. Hereafter, we refer to the crystalline water-ice feature that scales with the extinction opacity as the \textit{extinction feature}.

\subsection{Fresnel feature evaluation}
Even in PPDs, a 3.1~\mum\ feature is visible in their reflection spectra, although it is generally weaker than simulated in debris disks. The  lower right panel of Figure~\ref{fig:PPDs_spectra} compares the analytic relation \eqref{eq:fitting_delta_fresnel} derived from debris-disk spectra with the $c$--$\Delta_{\rm Fresnel}$ relation measured for PPDs. For the edge-on case, we quantify the depth of the additional absorption at 3.1~\mum\ and adopt its absolute value as $\Delta_{\rm Fresnel}$ (see the lower left and middle panels of Figure ~\ref{fig:PPDs_spectra} for how the pseudo-continuum is drawn). Using this $\Delta_{\rm Fresnel}$ metric, we find that PPDs also exhibit an approximately linear relation between crystallinity and $\Delta_{\rm Fresnel}$, similar to debris disks.

However, although we assume $f_{\rm rock}=1$ in regions with $T<150$~K, $\Delta_{\rm Fresnel}$ is lower at almost all crystallinities than in the rock-rich debris-disk case ($f_{\rm rock}=3$). This reduction may reflect both the fact that the scattered-light spectrum of a protoplanetary disk scales with $\omega$, whose Fresnel feature is weaker than that of $\kappa_{\rm sca}$ (Figure \ref{fig:optical_property_Mie}), and contributions to the disk flux from relatively warm ice in the inner disk (Appendix \ref{append:temperature_effect}). Therefore, the non-detection of a clear Fresnel feature in observed PPD spectra should not be taken as immediate evidence for the absence of crystalline ice. Conversely, integral-field spectroscopy that selectively probes scattered light from the outer disk and upper layers may enhance the Fresnel feature and improve detectability (see Section \ref{subsec:eppd_var}).

\subsection{Spatial variability of spectra in edge-on PPDs}
\label{subsec:eppd_var}
%%%%%%%%%%%%%%%%%%%%%%%%%%%%%%%%%%%
\begin{figure}[t]
\includegraphics[width=\hsize]{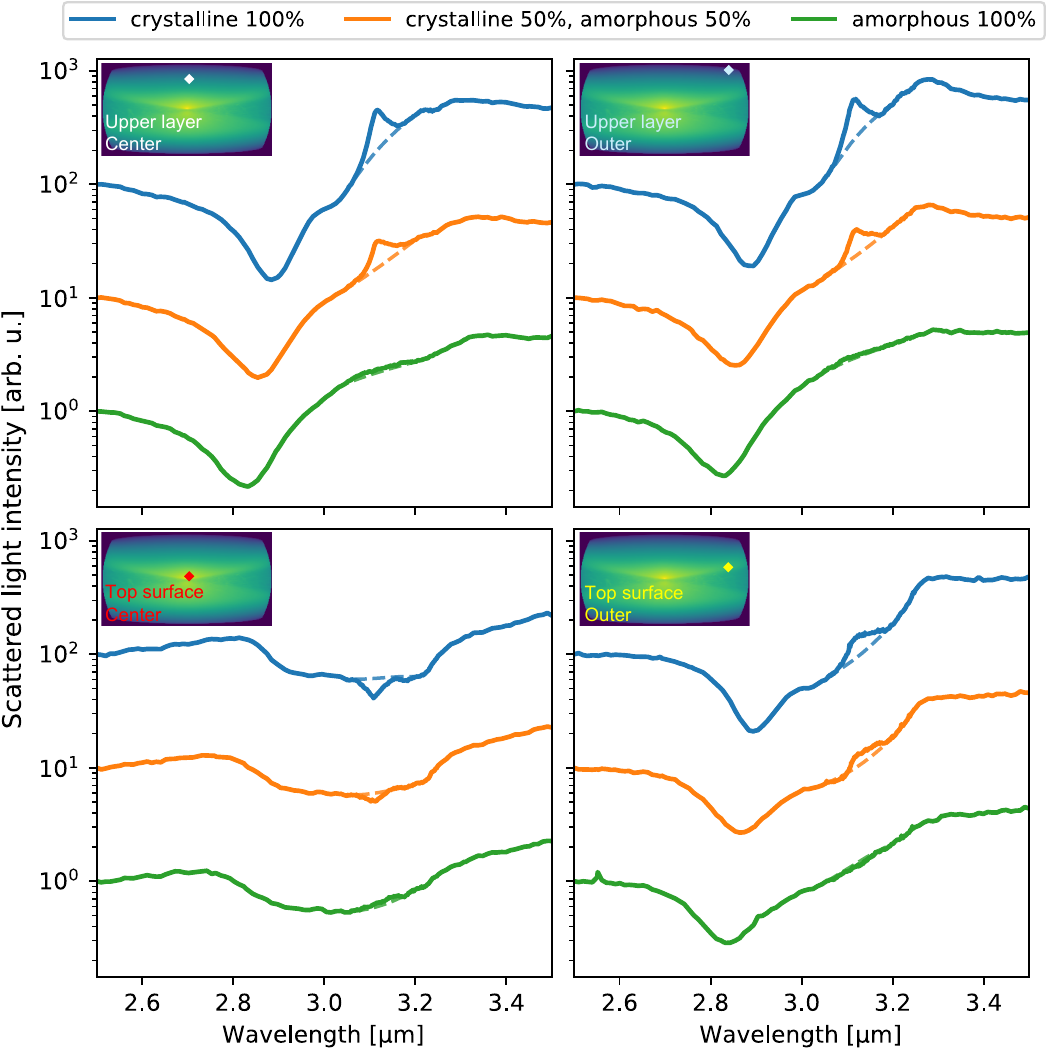}
\caption{Spatial variability of the transmitted- and scattered-light spectra in the edge-on PPD. 
In each panel, the corresponding extraction position is marked in the inset. 
The top row shows scattered-light spectra from the upper layers: the left panel is extracted near the image center, and the right panel is extracted toward the outer disk. 
The bottom row shows transmitted-light spectra along the top surface: the left panel is extracted near the image center, and the right panel is extracted toward the outer disk. {Alt text: Spectra of the edge-on PPD in different regions.}}
\label{fig:spatial_variability_eppd}
\end{figure}
%%%%%%%%%%%%%%%%%%%%%%%%%%%%%%%%%%

Edge-on PPD imaging is dominated in terms of the disk-integrated flux by transmitted light from the bright central region. In this geometry, the 3.1~\mum\ crystalline water-ice feature tends to appear as an extinction feature, whereas integral-field spectroscopy may reveal substantial spatial variability. Figure~\ref{fig:spatial_variability_eppd} presents spectra extracted from the midplane and the upper layers, each evaluated both near the image center and outer region of the image. In the top surface--center region, which contributes most strongly to the disk-integrated flux, a 3.1~\mum\ feature appears in absorption with a depth that varies with crystallinity. By contrast, in the top surface--outer region the spectrum develops an albedo feature reminiscent of the face-on case. In the upper layers, where the contribution from scattered light is larger, a Fresnel peak emerges, as in debris disks. These results suggest that spatially resolved spectroscopy with JWST could, in principle, constrain the spatial diversity of ice crystallinity by measuring the feature heights in the upper layers and the midplane and comparing them with our models.

However, JWST observations of edge-on disks have reported that even integral-field data can yield similarly saturated spectra across the field, potentially washing out intrinsic spatial variations \citep{2023A&A...679A.138S}. Multiple scattering has been suggested as a plausible explanation for this behavior. We therefore performed additional radiative-transfer calculations in which scattering was restricted to single scattering. The disk temperature structure, geometrical structure, and grain-size distribution were the same as those of the face-on and edge-on disk models examined in Section~\ref{subsec:ppd_spectra}. We used $5\times10^7$ photon packets for imaging and verified that the resulting spectra were sufficiently smooth, without noticeable Monte Carlo noise. Nevertheless, we found no substantial change in the overall spectral morphology compared to our fiducial calculations that allow multiple scattering. Determining whether the spatial diversity of crystallinity can be robustly tested will require further investigation, informed by the limitations and systematics of JWST integral-field observations.

\section{Discussions}
\label{sec:discussions}
\subsection{Interpreting ice features: Trends from the parameter study}
Our simulations of scattered light from debris disks show that the near-infrared spectral morphology depends on the ice crystallinity, ice-grain size, rock-to-ice mass ratio, and disk geometry. For disks composed of ice-containing grains, all cases share the O--H stretching absorption feature near 2.9~\mum, but the subsequent brightening toward $\sim$3.4~\mum\ and the longer-wavelength spectral slope vary with grain size (Figure~\ref{fig:debri_spectra}). This wavelength range includes the 3.1~\mum\ Fresnel feature, a key diagnostic for distinguishing crystalline from amorphous ice. Notably, only the $a=1~\mu\mathrm{m}$ case places the Fresnel peak partly embedded in a rising continuum, whereas for the other grain sizes the Fresnel peak is clearly resolved as a distinct peak. The result that rock-rich spectra dominated by $a=1~\mu\mathrm{m}$ grains show only a kink-like Fresnel feature even at high crystallinity. In our $\Delta_{\rm Fresnel}$ measurement, kink-like Fresnel features can yield values of $\Delta_{\rm Fresnel}$ comparable to those of distinct peak-like features. However, to avoid overlooking signatures of crystalline ice in observed spectra with kink-like structures, it is also important to constrain the grain size using additional spectral diagnostics. In this regard, spectra dominated by $a=1~\mu\mathrm{m}$ grains can be distinguished from those for other grain sizes by the fact that the 3.25~\mum\ secondary peak has a larger amplitude than the 3.1~\mum\ Fresnel feature. In addition, the presence or absence of the 2.0~\mum\ absorption feature, which is characteristic of large ice grains, may provide an independent indicator of grain size (see Appendix~\ref{append:2um_feature}). A Fresnel feature embedded in a rising continuum resembles that observed in the debris disk HD~181327 \citep{2025Natur.641..608X}.

\begin{figure*}[t]
\includegraphics[width=\hsize]{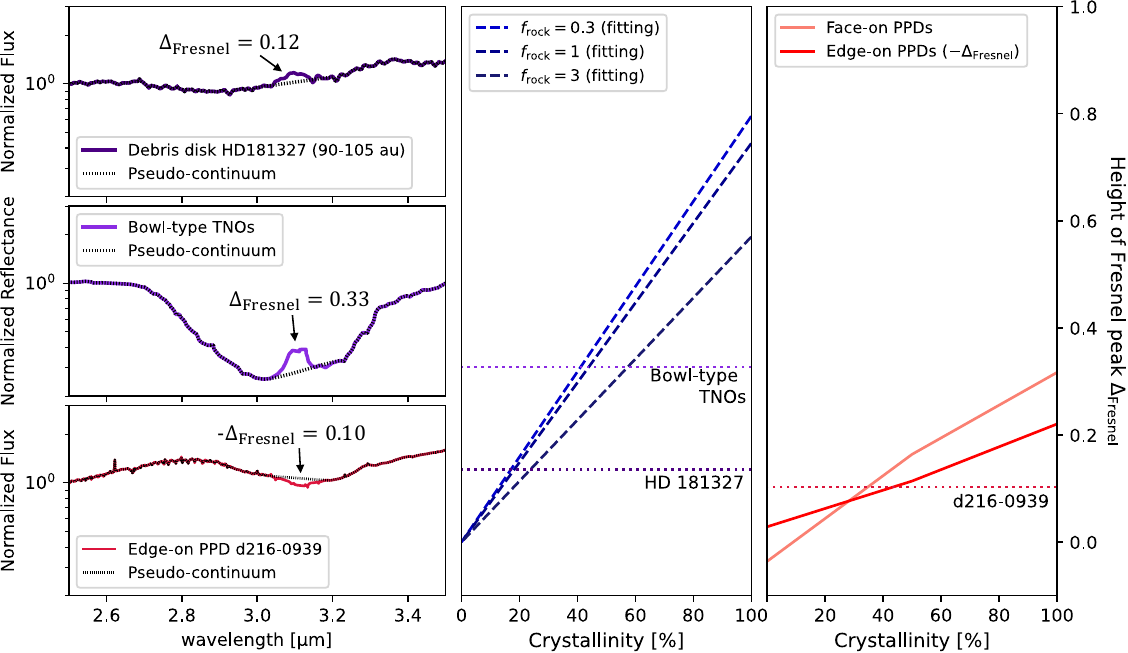}
\caption{Comparison of observed $\Delta_{\rm Fresnel}$ and the fitting function \eqref{eq:fitting_delta_fresnel}.
Left: Observed spectra (solid lines) of systems with a Fresnel peak detected by JWST, together with the pseudo-continuum constructed by masking the Fresnel peak (dashed lines). The debris disk HD 181327 (upper left) uses the public data of \citet{2025Natur.641..608X}. The mean of Bowl-type Trans-Neptunian Objects (middle left) and the protoplanetary disk d216-0939 are traced with WebPlotDigitizer from \citet{2025NatAs...9..245L} and \citet{2025A&A...697A..53P}, respectively.
Middle: The fitting function \eqref{eq:fitting_delta_fresnel}, identical to that in Figure \ref{fig:comparison_fit_and_model} (dashed lines). Dark-purple and light-purple dotted lines show $\Delta_{\rm Fresnel}$ estimated from the observed spectra for HD 181327 and for bowl-type TNOs, respectively.
Right: The crystallinity dependence of Fresnel-feature heights in PPDs, identical to that in Figure \ref{fig:PPDs_spectra} (red solid lines), with $\Delta_{\rm Fresnel}$ estimated from the observed spectra for d216-0939 from \citet{2025A&A...697A..53P}. {Alt text: Graphs showing the Fresnel feature height in observed systems and inferred crystallinity.}}
\label{fig:comparison_obs}
\end{figure*}

However, observations of outer Solar System bodies reveal a diversity of spectral shapes around 3~\mum, including "bowl-type" spectra that show an overall attenuation from 2.9 to 3.5~\mum\ and "cliff-type" spectra that dim near 2.9~\mum\ and do not brighten thereafter \citep{2025NatAs...9..230P, 2025NatAs...9..245L}. Such shapes are not produced by a face-on debris disk composed of water ice grains. In our simulations, a bowl-like profile is qualitatively reproduced when the rock-to-ice mass ratio is high (Figure~\ref{fig:debri_spectra}, second row), but it should be noted that our models do not include the 3.3~\mum\ C--H stretching absorption feature associated with organics or other carbon-bearing compounds. The presence of a C--H band could further enhance the prominence of the 3.1~\mum\ Fresnel peak; indeed, some TNOs exhibit a distinct Fresnel peak near the center of the bowl-shaped feature (see also Figure~\ref{fig:comparison_obs}). By contrast, we do not reproduce a cliff-type spectrum for any of the simulated combinations of grain size and rock-to-ice ratio. Therefore, a comparison between our simulations and cliff-type spectra requires careful consideration of which surface grain populations dominate the observed reflectance spectra of small bodies.

In addition, our protoplanetary disk simulations predict that the Fresnel feature appears in the disk-integrated flux of face-on disks as an albedo feature and in that of edge-on disks as an extinction feature (Figure \ref{fig:PPDs_spectra}). To our knowledge, this study provides the first theoretical investigation of the Fresnel feature expected in protoplanetary disks. It therefore offers a foundation for deriving the crystallinity of water ice from high-resolution spectra obtained with facilities such as JWST.

\subsection{Crystallinity of H$_2$O ice in the observed system}
\label{subsec:compared_obs}
%%%%%%%%%%%%%%%%%%%%%%%%%%%%%%%%%%%
%%%%%%%%%%%%%%%%%%%%%%%%%%%%%%%%%%

Determining the crystallinity of water ice in Solar System and extrasolar objects provides constraints on the heating and irradiation events experienced by the constituent grains. In this section, we focus on the Fresnel peak in observational data, estimate the ice crystallinity, and discuss the thermal history of the system’s dust grains. 

\subsubsection{Debris Disks and TNOs: Remnants of Planet Formation}
\label{subsubsec:compared_debris}
At present, a clear 3.1 \mum ~Fresnel peak has been reported from JWST/NIRSpec observations for the debris disk HD 181327 \citep{2025Natur.641..608X} and for Trans-Neptunian Objects that exhibit a bowl-shaped 3 \mum ~H$_2$O feature ("bowl-type" TNOs,  \citealp{2025NatAs...9..245L, 2025NatAs...9..230P}). The left panels of Figure \ref{fig:comparison_obs} show the observed spectra of these objects, the continua constructed by masking the Fresnel peak, and the corresponding values of $\Delta_{\rm Fresnel}$. The middle panel of Figure \ref{fig:comparison_obs} compares the fitting function presented in Figure \ref{fig:comparison_fit_and_model} with the $\Delta_{\rm Fresnel}$ values derived for these objects.

JWST integral-field spectroscopy of HD 181327 resolved three disk regions at different radial distances. Water ice is undetected in the Middle ring (80--90 au), detected in the Outer ring (90--105 au) and the Halo (105--120 au), and the Fresnel peak appears only in the Outer ring. The shape of the scattered light spectrum qualitatively resembles our debris disk models with a rock-rich composition ($f_{\rm rock} = 3$) and grain sizes $\gtrsim 1$ \mum, consistent with the rock-to-ice mass ratios and grain sizes inferred by \citet{2025Natur.641..608X}. From the spectrum we measure $\Delta_{\rm Fresnel} = 0.12$, implying an ice crystallinity of 10--20\%. Although HD 181327 has an inclination of $i=30^\circ$, its disk spectrum has been extracted from a region where the scattering angle is approximately $90^\circ$. Therefore, it is reasonable to estimate the crystallinity using equation \eqref{eq:fitting_delta_fresnel}, which is intended for $90^\circ$ scattering.

The small but nonzero crystalline fraction suggests that the grains had been heated to 140 K in their parent bodies, after which they were released recently. This interpretation may be supported by the expectation of sufficiently strong UV irradiation in this disk to photodesorb H$_2$O, and by polarimetric detections of sub-micron grains smaller than the blowout size \citep{2024A&A...683A..22M}. The absence of crystallinity signatures in the Halo region could simply indicate amorphous ice condensed in a colder environment; alternatively, the Halo grains may have been released earlier than those in the Outer ring and transported outward by blowout, undergoing UV-induced re-amorphization. While interpreting the integral-field data from the standpoint of grain migration is important, such modeling is beyond the scope of this study.

The mean reflectance spectrum of bowl-type TNOs exhibits a clearer Fresnel peak than that of HD 181327, and the inferred crystallinity ranges from 40--60\% depending on the rock-to-ice mass ratio. For bowl-type TNOs, the rock-to-ice mass ratio has been estimated to be $\sim$1 for objects such as Thereus that show deep CO$_2$ features, and $\gtrsim 4$ for objects such as Okyrhoe that show shallow CO$_2$ features \citep{2025NatAs...9..245L}. Resurfacing processes for TNOs include cosmic-ray irradiation, collisions, and enhanced insolation near perihelion \citep{2002P&SS...50...57G, 2009AJ....137.4296J}. Irradiation alters surface composition by promoting the production of complex organics and by re-amorphizing water ice, whereas collisions can erode irradiated surface layers and expose interior material. If internal heating events had crystallized H$_2$O ice within TNOs, collision-driven resurfacing may expose crystalline ice. However, the third process likely contributes most strongly to crystallization. Indeed, comet-like activity has been observed in some Centaurs beyond the water snowline, interpreted as a phase transition of amorphous ice driven by increased insolation, releasing volatile molecules previously trapped within the ice. In the Solar System, this transition becomes efficient at $\sim 10$ au, so the high H$_2$O crystallinities inferred for TNOs likely indicate ice that has undergone this transition \citep{2012AJ....144...97G}.

\subsubsection{Protoplanetary disks: Building Phase of planets }
\label{subsubsec:compared_PPD}
While JWST observations of debris disks and outer Solar System bodies are beginning to reveal signatures of crystalline ice, no clear near-infrared crystalline-ice feature has yet been reported in protoplanetary disks. \citet{2023A&A...679A.138S} suggested that the spectrum of the edge-on protoplanetary disk HH~48~NE may contain crystalline water ice; however, it exhibits neither the Fresnel peak seen in the debris disk HD181327 and TNOs nor additional absorption feature at 3.1 \mum\ predicted in our simulation. Prior to the advent of JWST, near-infrared observations with the Subaru Telescope had already suggested that some edge-on disks, such as PDS~453 and d216--0939, may contain highly crystallized water ice \citep{2012ApJ...753...19T, 2017ApJ...834..115T}. Among these, d216--0939 has recently been observed with JWST by \citet{2025A&A...697A..53P}. While that study primarily focused on identifying a variety of ice species, including NH$_4^+$ salts, the published spectrum appears to show a absorption near 3.1~\mum.  Thus we constructed a pseudo-continuum for d216--0939, measured $-\Delta_{\rm Fresnel}$ at 3.1~\mum, and compared the result with our PPD simulations (Figure~\ref{fig:comparison_obs}). We obtain $-\Delta_{\rm Fresnel}=0.1$ for d216--0939, smaller than the values reported so far for systems with a Fresnel peak. However, our models predict that, in protoplanetary disks, $\Delta_{\rm Fresnel}$ is systematically lower than in debris disks at any given crystallinity. Therefore, the inferred crystallinity for d216--0939 is relatively high, $\sim 50\%$. This result suggests that crystalline ice may be abundant even in systems at the planet-formation stage. 
This may suggest that ice thermally processed near the snowline, where temperatures approach the $\sim$140~K phase transition from amorphous to crystalline water ice, has been efficiently mixed throughout the disk. Protoplanetary disks exhibit processes such as turbulence and disk winds that can mix material radially and vertically, potentially transporting thermally processed ice from the inner disk to the disk surface in the outer disk \citep[e.g., ][]{1996Sci...271.1545S, 2001A&A...378..192G, 2007Sci...318..613C}. Indeed, in the Solar System, crystalline silicates that formed in the hot inner protosolar nebula are observed in comets, implying efficient large-scale mixing during planet formation \citep{2011AJ....141...26H, 2011AJ....142...80S, 2018AJ....156..242S}. 

On the other hand, it remains unclear whether PPDs ubiquitously contain crystalline ice. Some edge-on disks observed with JWST, such as HH 48 NE \citep{2023A&A...679A.138S}, show water-ice features but lack a clear 3.1~\mum\ Fresnel feature. We also applied our 3.1~\mum\ feature-extraction procedure to the Subaru spectrum of PDS~453 in \citet{2017ApJ...834..115T} but obtained $\Delta_{\rm Fresnel}=0$. However, because $\Delta_{\rm Fresnel}$ in PPDs are expected to be smaller than in debris disks, it is premature to interpret the absence of a 3.1~\mum\ feature in earlier observations as evidence for the absence of crystalline ice. Establishing how common crystalline ice is in PPDs will require a broader sample of disks observed at high spectral resolution with facilities such as JWST, together with complementary constraints from the far-infrared crystalline H$_2$O features at 44 and 62~\mum\ that are expected to become accessible with PRIMA \citep{2023prim.book.....M}.

\subsection{Potential mechanisms for ice crystallization}
While signatures of crystalline ice have been reported in a variety of systems spanning a wide range of ages and temperatures, it remains unclear how crystallization proceeded in each case. The amorphous-to-crystalline phase transition temperature of interstellar water ice is typically $\sim$140~K. Because this temperature is close to the sublimation temperature of water ice ($\sim$160~K), only a small fraction of icy dust in a given system is expected to reside in regions where crystallization can occur efficiently. However, this temperature constraint may be relaxed if the ice is allowed to crystallize over long timescales. The crystallization temperature $T_{\rm cryst}$ can be expressed as a function of the available time $t$ \citep{1996ApJ...473.1104J, 2016A&A...593A..11M}:
\begin{equation}
T_{\rm cryst}=\frac{T_{\rm lab}}{(R T_{\rm lab}/E_a)\,\ln(t/t_{\rm lab})+1},
\end{equation}
where $T_{\rm lab}=140$~K is the laboratory crystallization temperature, $t_{\rm lab}=1/365$~yr is the laboratory timescale, $R=8.314\times10^{-3}$~kJ~mol$^{-1}$~K$^{-1}$ is the gas constant, and $E_a=59.2$~kJ~mol$^{-1}$ is the activation energy for crystallization.

If disk grains can typically afford crystallization times comparable to orbital periods of $\sim$1--100~yr, the required temperature decreases to $\sim$120~K. Moreover, if the ice in TNOs is primordial and has survived since the protosolar nebula, it has had gigayear timescales available; in that case, the crystallization temperature can be reduced to $\lesssim$100~K. Thus, the high crystallinities inferred for some TNOs may reflect long-term annealing in environments with temperatures of order $\sim$100~K. Even so, in planet-forming disks the region where crystallization can occur efficiently without additional heating remains spatially limited. Without an efficient process that continuously mixes dust from the crystallization zone to the outer disk, achieving high crystallinities may require additional heating mechanisms.

In PPDs, episodic outbursts as observed in FU~Ori-type systems could transiently heat large regions of the disk \citep[e.g., ][]{1996ARA&A..34..207H, 2014prpl.conf..387A, 2016Natur.535..258C}. From far-infrared observations of HD~141527, \citet{2016A&A...593A..11M} argued for a high ice crystallinity and showed that an accretion burst with a rate of $\sim10^{-4}\,M_\odot\,{\rm yr}^{-1}$ could induce crystallization. In addition, shock heating driven by giant planets may also act as a transient disk-heating mechanism \citep{2020A&A...633A..29Z, 2025PASJ...77..149O, 2026PASJ...78..673O}. For more evolved systems that have undergone planetesimal and planet formation, such as debris disks, radiogenic heating from the decay of short-lived radionuclides inside parent bodies could provide an internal heat source \citep{2018SSRv..214...39M}. Revealing the origin of ice crystallinity will require both (i) expanding the statistical sample of crystallinity measurements across diverse environments and (ii) coupling these observations with detailed modeling of the thermal processes that can act on icy grains.

\subsection{Model caveats}
\label{subsec:caveats}
We estimated the crystallinity of water ice in actual systems based on theoretical scattered-light and transmitted-light imaging of debris disks and protoplanetary disks. However, the temperature and geometric structure of the target system can affect the inferred crystallinity. Here we summarize several caveats of our modeling and discuss their possible impact on the estimated crystallinity.

Our empirical relation \eqref{eq:fitting_delta_fresnel} is best suited to scattered light at a scattering angle of $90^\circ$ from cold ($\sim$50~K) debris disks and analogues bodies under similar conditions. Because increasing temperature acts to reduce the slope of the linear $c$--$\Delta_{\rm Fresnel}$ relation, the crystallinity inferred by applying equation~\eqref{eq:fitting_delta_fresnel} to an observed $\Delta_{\rm Fresnel}$ from a warmer disk should be regarded as a lower limit. For example, if $\Delta_{\rm Fresnel}=0.3$ is obtained for a face-on debris disk with a dust temperature of $\sim$100~K, equation~\eqref{eq:fitting_delta_fresnel} may underestimate the crystallinity by about 20\% (see Appendix~\ref{append:temperature_effect}). In addition, the Fresnel feature is most prominent in back-scattered light and becomes weaker toward forward-scattered light. Therefore, caution is required when comparing equation~\eqref{eq:fitting_delta_fresnel} with spectra extracted from regions where the scattering angle differs from $90^\circ$. In general, because the disk integrated flux of an inclined disk is dominated by bright forward-scattered light, applying equation~\eqref{eq:fitting_delta_fresnel} to such data would tend to underestimate the crystallinity. Similarly, although we estimated the crystallinity of bowl-type TNOs to be 40--60\% using equation~\eqref{eq:fitting_delta_fresnel}, it should be noted that this estimate does not incorporate information on the actual scattering angles of the observed objects.
As a rough estimate, we use our inclined debris-disk simulations to assess how deviations from $90^{\circ}$ scattering could affect the inferred crystallinity. For $\Delta_{\rm Fresnel}=0.33$ and $f_{\rm rock}=1$, if the observations of bowl-type TNOs include forward-scattered light, the inferred crystallinity would exceed 40\%. By analogy with the disk-integrated flux case (Figure~\ref{fig:Fresnel_params}, lower left), the inferred value would be 50\%. When we separately computed the $c$--$\Delta_{\rm Fresnel}$ relation using spectra extracted only from the bright forward-scattered light along the minor axis of the debris disk (scattering angle of $45^{\circ}$), the inferred crystallinity was as high as 80\%. By contrast, if the observations probe backscattered light, the inferred crystallinity would fall below 40\%. Using spectra extracted only from the faint backscattered light along the minor axis of the debris disk (scattering angle of $135^{\circ}$), the inferred value was 30\%. Therefore, the effect of phase angle on the crystallinity estimate for bowl-type TNOs is likely to be of order $\sim$10\%, although it could be substantially larger if the observed scattering angle is very small.

It should also be noted that equation~\eqref{eq:fitting_delta_fresnel} is based on simulations of optically thin debris disks, whereas it is not clear whether the scattered light obtained in TNO observations corresponds to that from an optically thin system. If the observed light instead reflects the scattering properties of an optically thick system, then the face-on PPD results would provide a more appropriate point of comparison for the $\Delta$ values derived from TNO spectra. In that case, the the inferred crystallinity of 40--60\% would represent a lower limit.

We also examined the dependence of $\Delta$ on crystallinity in protoplanetary disks, although our results are based on a specific disk structure and stellar properties. In addition, we fixed the rock-to-ice mass ratio at $f_{\rm rock}=1$ in regions with $T<150$~K, whereas actual disks may be more rock-rich and may also exhibit spatial variations in $f_{\rm rock}$. A comprehensive exploration of the effects of disk/stellar properties and composition on the Fresnel feature remains a subject for future work.

Finally, our estimates of crystallinity based on the observed spectra may include uncertainties introduced by digitization. Although WebPlotDigitizer has been reported to accurately recover y-axis values \citep{turner2023effect}, the 3.1 \mum\ feature in protoplanetary disks is intrinsically weak, and even small errors may affect the inferred crystallinity. If we conservatively assume a 5\% digitization-induced uncertainty in the flux, the value of $\Delta_{\rm Fresnel}$ obtained for d216-0939 would vary by approximately 0.05, corresponding to an effect of about 20\% on the estimated crystallinity.

\section{Conclusions}
\label{sec:conclusions}
In this study, we investigated how the Fresnel feature appearing in the near- to mid-infrared spectra of debris disks and protoplanetary disks (PPDs) varies with icy dust properties and disk geometry, with the aim of providing a framework for estimating ice crystallinity from this feature. Our key findings are as follows.

\begin{enumerate}
 \item We found a universal relation that the scattered-light spectra of debris disks containing crystalline water ice, when extracted around a scattering angle of $90^\circ$, exhibit an upward peak- or kink-like Fresnel feature at 3.1~\mum, whose height, $\Delta_{\rm Fresnel}$  (defined as the logarithm of the ratio of the spectral intensity at 3.1~\mum\ with the Fresnel feature to that without it) increases linearly with crystallinity. This relation holds regardless of the characteristic grain size of the disk.

 \item In face-on PPDs, the 3.1~\mum\ Fresnel feature appears as an albedo feature that becomes more pronounced with increasing crystallinity (Figure \ref{fig:PPDs_spectra}). In edge-on PPDs, the disk-integrated flux is dominated by transmitted light from the central region of the image, and the 3.1~\mum\ crystalline water-ice feature appears as an extinction feature. When the scattered-light component from the upper layers of the edge-on disk is selectively extracted, the 3.1~\mum\ Fresnel feature appears as a peak or kink in the spectrum.

 \item The Fresnel-feature height $\Delta_{\rm Fresnel}$ in PPDs is smaller than that in debris disks for any crystallinity (Figure \ref{fig:PPDs_spectra}). We have shown that this is not due to multiple scattering, but is likely due either to the intrinsically weaker crystalline feature in $\kappa_{\rm ext}$ and $\omega$ than in $\kappa_{\rm sca}$, or to contributions from relatively warm dust in the inner regions of the disk.

 \item Applying our scaling relation (equation \ref{eq:fitting_delta_fresnel}), we derived the crystallinity of water ice in the debris disk HD~181327, bowl-type TNOs, and the protoplanetary disk d216-0939 to be 10--20\%, 40--60\%, and $\sim$50\%, respectively (Figure \ref{fig:comparison_obs}).
 
 \item Although the crystallization mechanisms remain uncertain, the TNOs may have undergone long-term annealing at temperatures of $\sim 100$~K over timescales comparable to the age of the Solar System, whereas the protoplanetary disk may have experienced tentative heating events such as accretion bursts. In HD~181327, the low crystallinity may reflect re-amorphization driven by intense ultraviolet irradiation.
\end{enumerate}

Our simulations demonstrate that the Fresnel feature provides a useful diagnostic for quantitatively constraining the crystallinity of water ice in planet-forming disks. In this study, we estimated the ice crystallinity for one debris disk and one protoplanetary disk based on JWST data. As more such observations become available in the future, it should become possible to discuss the crystallinity of disk ice statistically, thereby helping to reveal the thermal and material-transport histories of planet formation.

\begin{ack}
 We are sincerely grateful to Dr. Minjae Kim for the constructive and insightful review of our manuscript. We would like to thank Olivier Poch, Hiroyuki Kurokawa and Satoshi Okuzumi for their helpful comments.
\end{ack}

\section*{Funding}
This work was supported by JSPS KAKENHI Grant numbers 25KJ0093 and 25K01049.

%%%%%%%%%%%%%%%%%%%%%%%%%%%%%%%%
\bibliographystyle{apj}
\bibliography{article} 
%%%%%%%%%%%%%%%%%%%%%%%%%%%%%%%%

\appendix

\section{Analytic expressions for optical properties of large, less absorbing grains}
\label{append:ADtheory}
Within the 2.9--3.3~\mum\ range that includes the Fresnel feature, the optical properties of ice grains obtained from Mie calculations are well reproduced by two limiting approximations: the Rayleigh-regime expressions for small grains ($x\ll 1$) and the geometric-optics expressions for large grains ($x\gg 1$). However, outside 2.9--3.3~\mum, the geometric-optics approximation no longer performs well. This is because, at these wavelengths, the imaginary part $k$ of ice becomes small and the grains are optically thin for both scattering and absorption (i.e., $|m-1|<1$ and $kx<3/8$).
For such large, transparent grains, alternative approximations that reproduce the Mie-calculated $Q_{\rm abs}$ and $Q_{\rm sca}$ include the geometric-optics approximation for less absorbing spheres and the Rayleigh--Gans approximation \citep{1983asls.book.....B}:
\begin{equation}
Q_{\rm abs}^{\rm GO(less\,abs)} = \frac{8kx}{3n}\left[n^{3}-\left(n^{2}-1\right)^{3/2}\right],
\end{equation}
\begin{equation}
Q_{\rm sca}^{\rm Rayleigh\text{-}Gans} = 2x^{2}|m-1|^{2}.
\end{equation}
A convenient prescription for large grains is \citep{2014A&A...568A..42K}
\begin{equation}
Q_{\rm abs} = \min\!\left(Q_{\rm abs}^{\rm GO(less\,abs)},\,Q_{\rm abs}^{\rm GO}\right),
\end{equation}
\begin{equation}
Q_{\rm sca} = \min\!\left(Q_{\rm sca}^{\rm Rayleigh\text{-}Gans},\,Q_{\rm sca}^{\rm GO}\right).
\end{equation}
We verified that this prescription reproduces the Mie-calculated $Q_{\rm abs}$ for large grains over 2.5--3.3~\mum, whereas the agreement for $Q_{\rm sca}$ is poorer.

Another approximation applicable to large, transparent particles is anomalous diffraction theory \citep{1957lssp.book.....V}:
\begin{equation}
\begin{split}
    Q_{\rm ext}^{\rm AD} = &2
-4\exp(-\rho\tan\beta)\frac{\cos\beta}{\rho}\sin(\rho-\beta)
\\&-4\exp(-\rho\tan\beta)\left(\frac{\cos\beta}{\rho}\right)^{2}\cos(\rho-2\beta)
\\&+4\left(\frac{\cos\beta}{\rho}\right)^{2}\cos 2\beta,
\end{split}
\end{equation}

\begin{equation}
Q_{\rm abs}^{\rm AD} = 1 + \frac{\exp(-2\rho\tan\beta)}{\rho\tan\beta}
+ \frac{\exp(-2\rho\tan\beta)-1}{2\rho^{2}\tan^{2}\beta},
\end{equation}

\begin{equation}
Q_{\rm sca}^{\rm AD} = Q_{\rm ext}^{\rm AD} - Q_{\rm abs}^{\rm AD}.
\end{equation}
Here, $\rho = 2x|m-1|$ and $\tan\beta = \mathrm{Im}(m)/\mathrm{Re}(m-1).$
From these, we obtain
\begin{equation}
Q_{\rm abs} = \min\!\left(Q_{\rm abs}^{\rm AD},\,Q_{\rm abs}^{\rm GO}\right).
\label{eq:Qabs_min_AD_GO}
\end{equation}
We use equation~\eqref{eq:Qabs_min_AD_GO} to select whether the AD or GO regime dominates. We confirmed that, by computing $Q_{\rm abs}$ and $Q_{\rm sca}$ with the corresponding AD or GO expressions in each regime, the Mie results for large ice grains can be reproduced (Figure~\ref{fig:optical_property_Mie}).

\section{Constructing Pseudo-Continuum}
\label{append:continuum_const}
%%%%%%%%%%%%%%%%%%%%%%%%%%%%%%%%%%%
\begin{figure}[t]
\includegraphics[width=0.8\hsize]{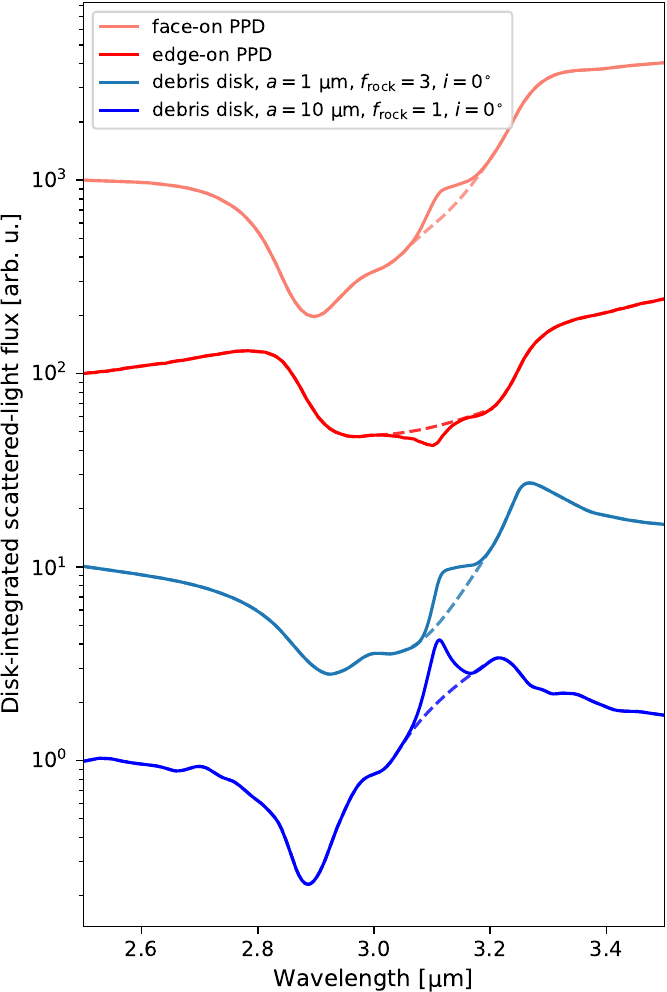}
\caption{Examples of pseudo-continua constructed to quantify the Fresnel feature. 
Debris-disk cases: the fiducial model with $a=10$~\mum\ (blue) and the rock-rich model with $a=1$~\mum\ (light blue), for which the 3.05--3.2~\mum\ interval is masked and interpolated with a cubic-spline function. 
Protoplanetary-disk case: the edge-on model (red) and the face-on model (light red), for which the 3.05--3.2~\mum\ interval is masked and interpolated using Akima interpolation. {Alt text: A graph showing the disk spectra and pseudo-continua.}}
\label{fig:pseudo-continuum}
\end{figure}
%%%%%%%%%%%%%%%%%%%%%%%%%%%%%%%%%%

Here we describe the interpolation methods used to construct the pseudo-continuum for evaluating crystalline-ice features and present several examples of the resulting pseudo-continua. All interpolations were performed using the Python package \texttt{SciPy} \citep{2020NatMe..17..261V}. After exploring several approaches, we found that, for interpolating the Fresnel feature in debris-disk spectra produced by \texttt{DDiT}, a cubic-spline interpolation with a masking range of 3.05--3.2~\mum\ yields a smooth continuum even for disks with different inclinations and high rock-to-ice mass ratios (Figure~\ref{fig:pseudo-continuum}; blue solid lines). A cubic spline is also effective for the secondary peak, but it requires spectrum-by-spectrum tuning of the masking wavelengths. By contrast, the PPD spectra exhibit low-level noise introduced by the Monte Carlo radiative-transfer calculation, for which spline interpolation can become unstable. Instead, we find that Akima interpolation provides a smooth pseudo-continuum across the Fresnel feature (Figure~\ref{fig:pseudo-continuum}; red solid lines). Akima interpolation is also effective for constructing pseudo-continua for noisier spectra such as observational data.

\section{Effect of ice temperature on Fresnel feature}
\label{append:temperature_effect}

%%%%%%%%%%%%%%%%%%%%%%%%%%%%%%%%%%%
\begin{figure}[t]
\includegraphics[width=\hsize]{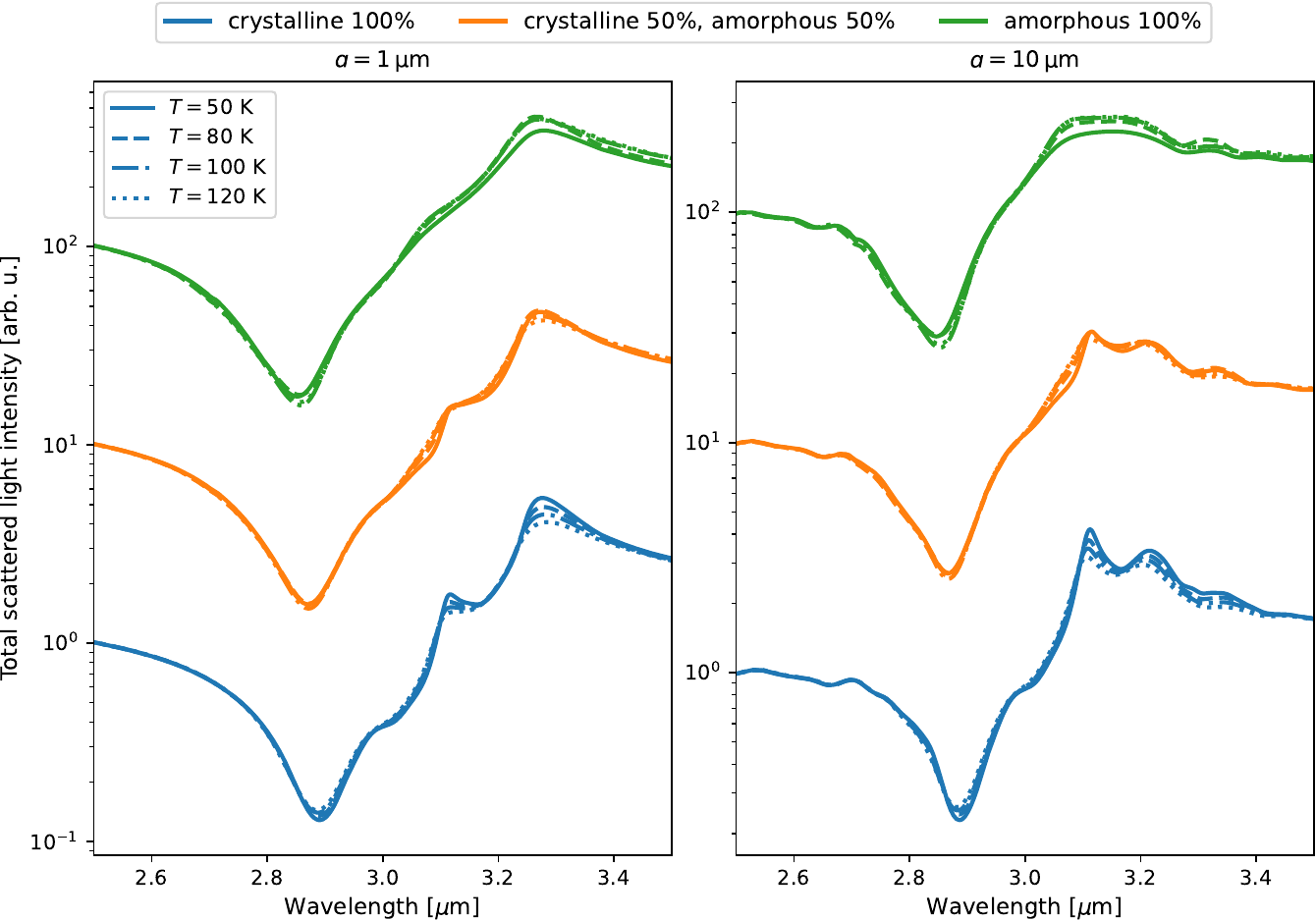}
\caption{Same as Figure~\ref{fig:debri_spectra}, but for different ice temperatures. The solid, dashed, dash--dotted, and dotted curves use the complex refractive indices of ice measured at 50, 80, 100, and 120~K, respectively. {Alt text: Two graphs showing the spectra for different ice temperatures.}}
\label{fig:spectra_temp}
\end{figure}
%%%%%%%%%%%%%%%%%%%%%%%%%%%%%%%%%%

%%%%%%%%%%%%%%%%%%%%%%%%%%%%%%%%%%%
\begin{figure}[t]
\includegraphics[width=\hsize]{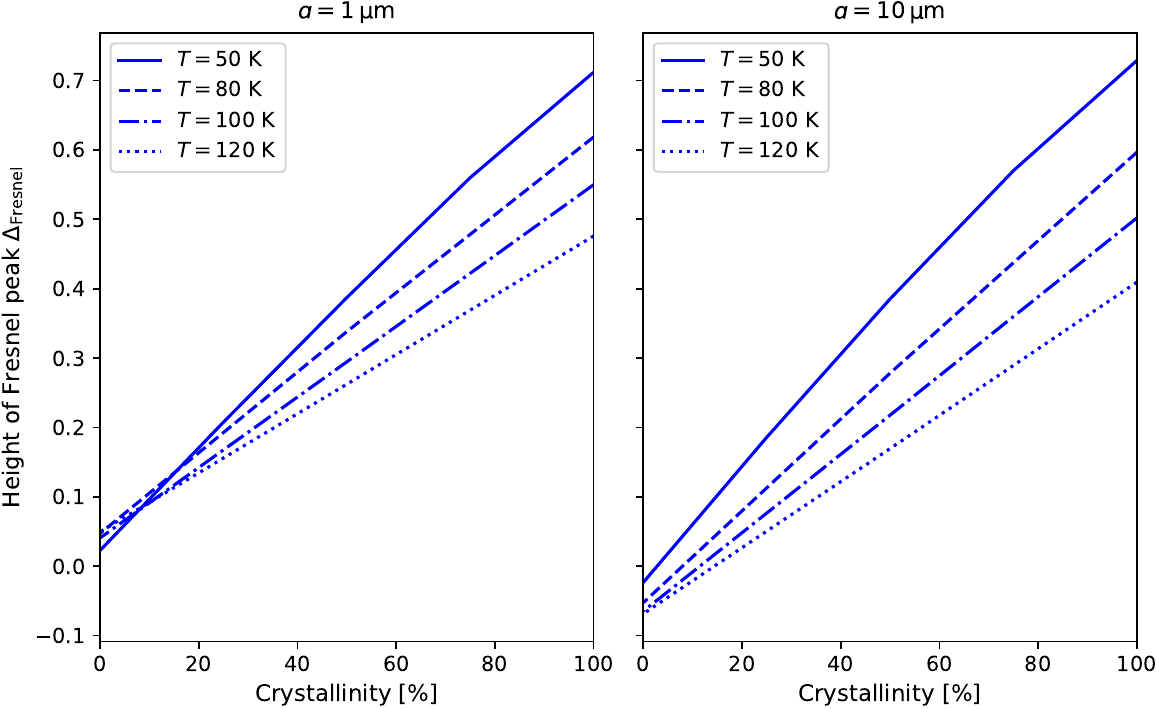}
\caption{Crystallinity dependence of the Fresnel-peak height for different ice temperatures. {Alt text: Two line graphs of $\Delta_{\rm Fresnel}$-$c$ relation for different ice temperatures.}}
\label{fig:Fresnel_temp}
\end{figure}
%%%%%%%%%%%%%%%%%%%%%%%%%%%%%%%%%%

In this study, we investigated how the Fresnel feature depends on grain size, the rock-to-ice mass ratio, and disk geometry. We compared our results with observations using laboratory-measured complex refractive indices of water ice at 50~K. This choice is appropriate for comparisons with debris disks and small bodies on orbits far beyond the water snowline. However, the complex refractive index of water ice is temperature dependent, which can in turn modify the disk scattered-light spectrum, including the Fresnel feature. In this section, we examine the temperature dependence of the scattered-light spectra of a face-on debris disk composed of pure-ice grains.

Figure~\ref{fig:spectra_temp} shows the predicted scattered light spectra of a face-on debris disk obtained using complex refractive indices measured at higher temperatures (80--120~K) than in our fiducial case. For all grain sizes, the overall spectral shape depends only weakly on ice temperature. In contrast, the depth of the O--H stretching feature near 2.9~\mum\ and the heights of the Fresnel and secondary peaks vary slightly with temperature. In particular, at high crystallinity, both peak heights decrease as the ice temperature increases.

To quantify the crystallinity dependence of the Fresnel peak height $\Delta_{\rm Fresnel}$ at each temperature, we construct the continuum in the same manner as in the fiducial case by masking 3.05--3.2~\mum\ and applying a spline interpolation, and then measure the peak height (Figure~\ref{fig:Fresnel_temp}). For all cases with $T\ge 80$~K, the dependence of $\Delta_{\rm Fresnel}$ on crystallinity is qualitatively similar across temperature. At high crystallinity ($c>0.5$), $\Delta$ is lower at higher temperatures. At low crystallinity ($c<0.5$), the temperature dependence is weak, although $\Delta_{\rm Fresnel}$ increases slightly with temperature. Therefore, if a large value such as $\Delta_{\rm Fresnel}>0.6$ is observed and crystallinity is inferred by comparison with our empirical relation, the inferred value should be regarded as a lower limit.

The stellar luminosities of HD~181327, for which we estimated the ice crystallinity using our models in Section~\ref{subsubsec:compared_PPD}, is $0.44\,L_\odot$. For an optically thin disk, the midplane temperature can be written as
\begin{equation}
\label{eq:temp_thin}
T \sim 280 \left(\frac{r}{1\,{\rm au}}\right)^{-1/2}\left(\frac{L}{L_\odot}\right)^{1/4},
\end{equation}
which gives $T\sim25$~K at $r=100$~au for HD~181327. Therefore, it is reasonable to estimate the crystallinity using our empirical relation \ref{eq:fitting_delta_fresnel} derived from the 50~K spectra.

\section{2 \mum\ absorption: A signature of large icy dust}
\label{append:2um_feature}
%%%%%%%%%%%%%%%%%%%%%%%%%%%%%%%%%%%
\begin{figure}[t]
\includegraphics[width=\hsize]{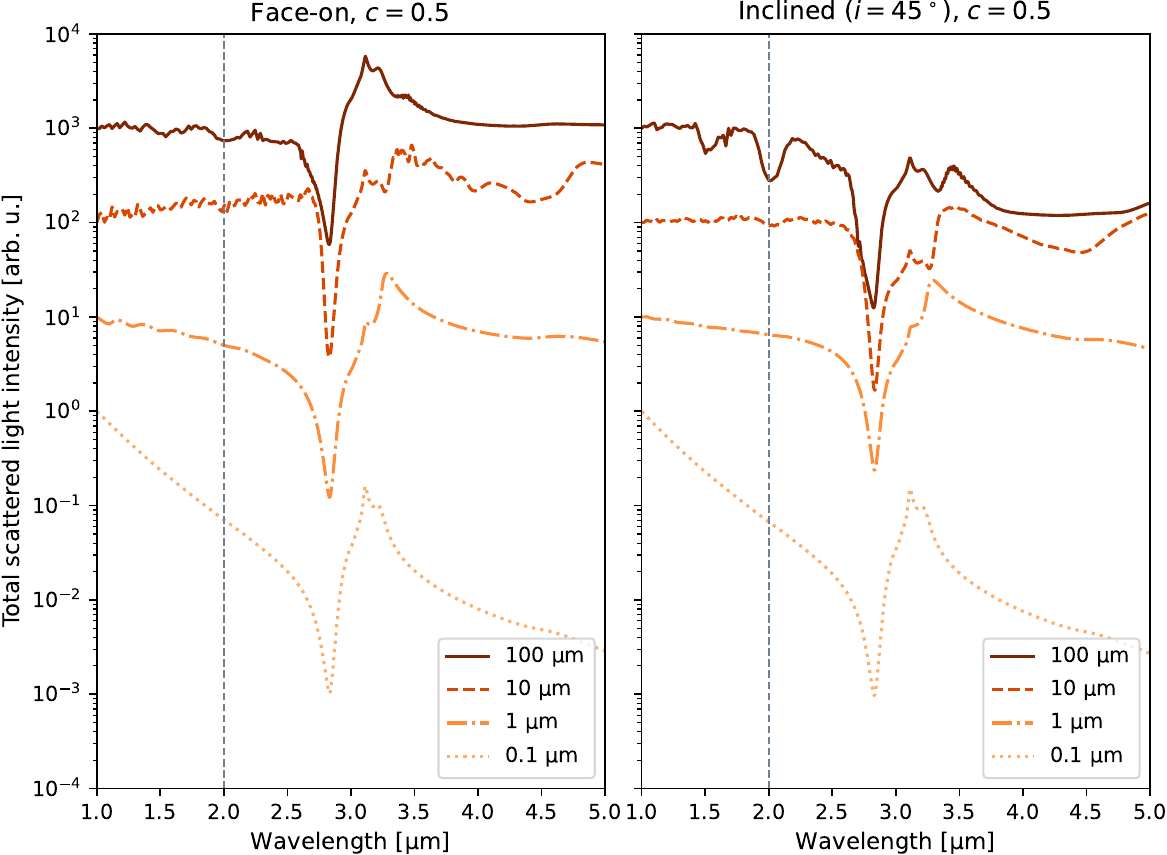}
\caption{Integrated scattered-light spectra of a debris disk composed of pure ice grains for different grain sizes. 
The left panel shows the face-on case, and the right panel shows the inclined case ($i=45^\circ$). 
The gray vertical lines highlight the $\sim$2~\mum\ H$_2$O absorption feature that appears in disks composed of grains with $a \protect\gtrsim 10~\rm{\mu m}$.
{Alt text: Extended integrated scattered-light spectra of a debris disks.}}
\label{fig:2um_absorb}
\end{figure}
%%%%%%%%%%%%%%%%%%%%%%%%%%%%%%%%%%

While our paper focuses on how the morphology and amplitude of the Fresnel feature near 3.1~\mum\ depend on model parameters, the H$_2$O absorption band near 2.0~\mum\ may provide an additional tracer of grain size and rock-to-ice mass ratio that is largely insensitive to crystallinity. Figure~\ref{fig:2um_absorb} extends the integrated scattered-light spectra of a debris disk composed of pure water ice ($c=0.5$) down to 1~\mum\ and compares the results for different grain sizes. The $\sim$2~\mum\ absorption feature emerges for $a=10$~\mum\ and becomes more prominent for $a=100$~\mum. These features are further enhanced when the disk is inclined; however, even in this geometry, they appear only for grains with $a\gtrsim10$~\mum. Moreover, the $\sim$2~\mum\ feature disappears when ice is mixed with more absorbing rocky material. In our models, this band is absent when we adopt \texttt{astrodust} for the complex refractive index of the rocky component, whereas it is visible at low rock-to-ice mass ratios when the rock component is the more transparent silicate mixture Mg$_{0.7}$Fe$_{0.3}$SiO$_3$ \citep{1995A&A...300..503D}. Therefore, the 2~\mum\ absorption feature may serve as evidence for the presence of large, nearly pure ice grains. In JWST observations, a $\sim$2~\mum\ absorption feature appears to be present in some bowl-type TNOs. This may indicate that large icy (nearly pure) grains reside on TNO surfaces and contribute to their reflectance spectra.

\end{document}